\documentclass[aps,pre,twocolumn,showpacs,superscriptaddress,10p,longbibliography]{revtex4-2}
\usepackage{amsmath}
\usepackage{amsfonts}
\usepackage{amsmath}
\usepackage{amssymb}
\usepackage[colorlinks=true,linkcolor=red]{hyperref}
\usepackage[sort&compress]{natbib}
\usepackage{graphicx}
\usepackage{color}
\usepackage{siunitx}

\newcommand{\bra}[1]{\ensuremath{\left\langle#1\right|}}
\newcommand{\ket}[1]{\ensuremath{\left|#1\right\rangle}}

\newcommand{\RNumb}[1]{\uppercase\expandafter{\romannumeral #1\relax}}

\begin{document}

\title{Quantum beats of a magnetic fluxon in a two-cell SQUID}

\author{I. N. Moskalenko}
\affiliation{National University of Science and Technology "MISIS", 119049 Moscow, Russia}
\affiliation{Russian Quantum Center, 143025 Skolkovo, Moscow, Russia}
\author{I. S. Besedin}
\affiliation{National University of Science and Technology "MISIS", 119049 Moscow, Russia}
\affiliation{Russian Quantum Center, 143025 Skolkovo, Moscow, Russia}
\author{S. S. Seidov}
\affiliation{National University of Science and Technology "MISIS", 119049 Moscow, Russia}
\author{M. V. Fistul}
\affiliation{National University of Science and Technology "MISIS", 119049 Moscow, Russia}
\affiliation{Russian Quantum Center, 143025 Skolkovo, Moscow, Russia}
\affiliation{Theoretische Physik III, Ruhr-Universit\"at Bochum, 44801 Bochum Germany} 
\author{A. V. Ustinov}
\affiliation{National University of Science and Technology "MISIS", 119049 Moscow, Russia}
\affiliation{Russian Quantum Center, 143025 Skolkovo, Moscow, Russia}
\affiliation{Physikalisches Institut, Karlsruhe Institute of Technology, 76131 Karlsruhe, Germany}

\date{\today}

\begin{abstract}
 We report a detailed theoretical study of a coherent macroscopic quantum-mechanical phenomenon - quantum beats of a single magnetic fluxon trapped in a two-cell SQUID of high kinetic inductance. We calculate numerically and analytically the low-lying energy levels of the fluxon, and explore their dependence on externally applied magnetic fields. The quantum dynamics of the fluxon shows quantum beats originating from its coherent quantum tunneling between the SQUID cells. We analyze the experimental setup based on a three-cell SQUID, allowing for time-resolved measurements of quantum beats of the fluxon.  
\end{abstract}
%\pacs{42.50.-p,74.81.Fa,74.50.+r}
\maketitle

\section{Introduction}
Long Josephson junctions and Josephson junction parallel arrays (JJPAs) are ideal experimental platforms to a study a dynamics of topological excitations, so-called \textit{magnetic fluxons}. These peculiar macroscopic objects are vortices of persistent current, each of them carrying one quantum of magnetic flux, $\Phi_0$. The classical dynamics of fluxons in the presence of dc and ac applied magnetic and electric fields has been studied in the past \cite{Ustinov-Review, Ustinov_Cirillo_Malomed}. A variety of fascinating phenomena, e.g. the relativistic effects \cite{Ustinov-Review}, bunching of fluxons \cite{bunch}, multisoliton excitations \cite{Multisoliton_Excitations}, the Cherenkov radiation of plasma oscillations \cite{Cherenkov1, Cherenkov}, nonequilibrium metastable states of fluxons \cite{Librstates}, escape from a potential well induced by thermal fluctuations \cite{Thermalfl}, just to name a few, have been observed. In spite of JJPAs being composed of many strongly interacting degrees of freedom, the classical dynamics of a large fluxon occupying multiple cells in the array can be precisely mapped to a  relativistic mechanics of its center of mass \cite{Ustinov-Review, Thermalfl}.

In superconducting quantum circuits and superconducting qubits such as single small Josephson junctions,  many-junctions SQUIDs etc., various coherent macroscopic quantum phenomena, e.g. macroscopic quantum tunneling, quantum beats, microwave induced Rabi oscillations, Ramsey fringes, have been observed
 \cite{Nakamura, Astafiev, Martinis, Rabi_oscillations}.
 %\cite{qubits1,qubits2,qubits3,qubits4}.
However, in this field, the quantum coherent dynamics of mobile excitations ("flying" qubits, not just microwave photons) remains unexplored. Here a natural question occurs: is it possible to observe the coherent quantum-mechanical effects in the dynamics of macroscopic topological excitations occurring in strongly interacting many-body quantum systems?

The first attempts to observe the quantum-mechanical dynamics with fluxons trapped in long Josephson junctions, were made almost twenty years ago. In Ref. \cite{fluxon} the macroscopic quantum tunneling and energy levels quantization have been observed for a single fluxon trapped in a potential formed in a continuous long Josephson junction. Moreover, more complicated incoherent quantum phenomena such as quantum oscillations and quantum dissociation of vortex-(anti)vortex pairs have been predicted and observed \cite{FistUstMalomed}. However, the coherent time-domain macroscopic quantum dynamics of fluxons has not yet been observed. It was realized that a main obstacle on this way is a relatively large size and thus low effective charging energy of fluxons in long junctions which, in turn, prevents maintaining its particle-like entity in the quantum regime.

\begin{figure}
	\includegraphics[width=1\columnwidth]{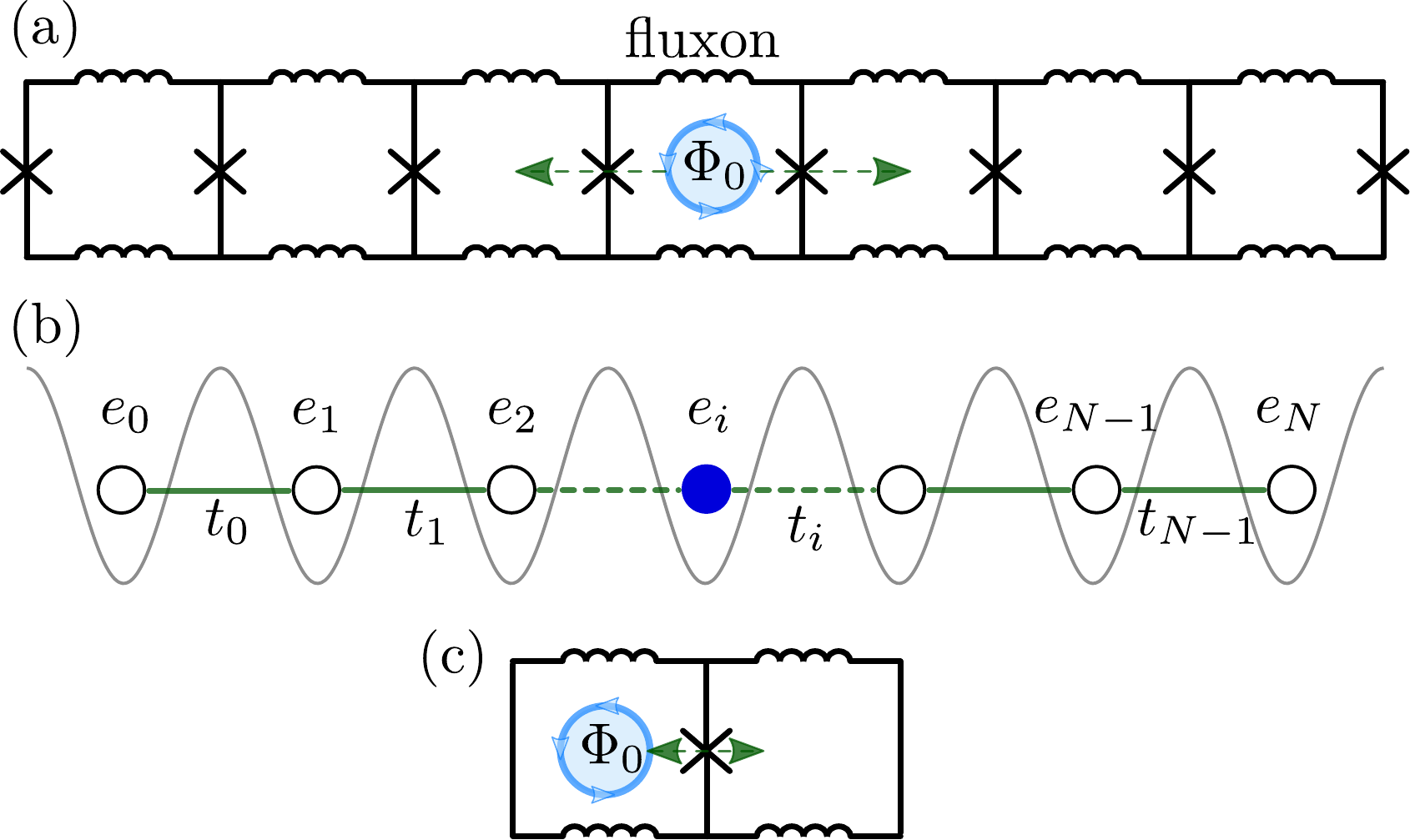}
\caption{(color online) (a) Schematic of a one-dimensional JJPA composed of many small Josephson junctions connected with inductors.  
(b) Periodic (Pierls-Nabarro) potential for a single fluxon trapped in the JJPA. Here, every site corresponds to a cell in (a), $e_{i}$ is on-site energy, $t_{i}$  is the tunneling rate. (c) The simplest primitive system - a two-cell SQUID for studying fluxon quantum dynamics. A trapped fluxon is shown.}
	\label{fig:JTL}
\end{figure}

The size of fluxons can be drastically reduced in JJPAs with high inductances of superconducting cells. JJPAs are composed of a chain of interconnected superconducting loops (cells) where the coupling between adjacent cells is provided by small Josephson junctions. The schematic of a such JJPA with a trapped fluxon is shown in Fig.~\ref{fig:JTL}a. The fluxon in JJPAs corresponds to a $2\pi$-kink in the distribution of phase drops across Josephson junctions, and the center of the stationary fluxon is positioned in one of the cells. 

Operation in the quantum regime requires very small Josephson junctions in JJPAs, small enough for their Josephson energy to be comparable to their charging energy. In this limit, an attempt to reduce the fluxon size to only few cells requires large  geometrical inductances, which makes the whole structure nonphysically large and strongly sensitive to parasitic fluctuations of the ambient magnetic fields. To avoid this problem, one can replace geometrical inductances by \textit{high kinetic inductances} by using chains of classical Josephson junctions \cite{MatveevGlazmanLarkin,Manucharyan} or disordered superconducting materials like granular aluminum (grAl) \cite{Maleeva,GranularAl}, NbTiN \cite{Nanowires}, or indium-oxide \cite{AstIoffePashkin}, similarly to the approach used for making fluxonium qubits.   
%    Figure 1

According to the theoretical analysis \cite{Petrescu}, the JJPAs with high kinetic inductances map to a wide class of 1D tight-binding lattice models and provide a quantum simulation platform for symmetry-protected topological excitations.
%with fractionally charged bound states at the edges.

The classical potential for a single topological kink in a sine-Gordon lattice is the so-called Pierls-Nabarro potential \cite{Pierls_Nabarro}. For a fluxon in an unbiased JJPA it is equivalent to $N$ degenerate minima separated by energy barriers centered at Josephson junctions. The larger is the size of magnetic fluxon compared to the array cell size, the lower is the height of these barriers (see Fig.~\ref{fig:JTL}b).
Hamiltonian for the one-fluxon manifold is:
 \begin{multline}
 \hat{H}_{tb}=\sum_{i=0}^{N-1} e_{i} \ket{i}\bra{i} + \sum_{i=0}^{N-2} t_{i}\ket{i}\bra{i+1},\ \ \ \ \ \ \ 
 \label{eq1}
 \end{multline} 
where $\ket{i}$ denotes the one-fluxon state at $i$-loop. In the Hamiltonian $\hat{H}_{tb}$ the interaction between nearest neighbors is taken into account, only. On-site energies $e_i$ are determined by the inductive energy of the cells and the Josephson energies of the junctions, and also are influenced by applied currents and magnetic fields penetrating the cells of JJPA. 
The amplitude $t_i$ of macroscopic quantum tunneling fo the fluxon between adjacent loops is determined by the ratio of Josephson coupling energy, $E_{Ji}$ and the charging energy $E_{Ci}$ as
$t_{i}\propto \exp{(-\sqrt{8E_{Ji}/E_{Ci}})}$ , see Appendix B for details. In such JJPAs, a spatial dependence of tunneling rates can be realized by modulating the Josephson coupling energies along the array.   

In this work, we theoretically study the coherent quantum dynamics of a fluxon trapped in a two-cell SQUID with high kinetic inductances (Fig.~\ref{fig:JTL}c). Such system presents the simplest setup for observing coherent quantum-mechanical 
dynamics of a single fluxon. Indeed, by manipulating external magnetic fields the fluxon can be trapped in one of the two cells (left or right). In the presence of substantial charging energy, $E_C$, one can expect the macroscopic quantum tunneling and coherent quantum beats of the fluxon between cells. In order to measure these (and other) quantum-mechanical effects in the dynamics of fluxons, we propose the following setup. The two cells of the SQUID are coupled via another superconducting cell containing two small Josephson junctions, thus forming a three-cell SQUID system (see, Fig.~\ref{fig:full_scematic}). In particular, the coherent quantum dynamics of a single fluxon trapped in the leftmost (rightmost) two cells can be measured by detecting the \textit{plasma modes} of the right (left) Josephson junction. In a way, this approach is similar to the conventional circuit cavity QED with a difference of using one of the Josephson junctions as the cavity.

The paper is organized as follows: In Section II we present a model for the three-cell SQUID of high kinetic inductance with two small Josephson junctions, derive the Lagrangian and the Hamiltonian of such system. In Section III by making use of numerical and analytical analysis we obtain the low-lying energy levels and energy level splitting of a fluxon trapped in the two-cell SQUID. The dependence of the quantum beats frequency on various parameters such as external magnetic fields, the kinetic inductances, and the Josephson coupling energy, is studied. In Sec. IV we extend our analysis to the three-cell SQUID and analyze its energy level structure. Furthermore, we suggest and explore the experimental protocol allowing one to measure of the energy levels and coherent quantum beats of the fluxon. Section V contains our conclusions. 

\section{Electromagnetic circuit, Lagrangian and Hamiltonian of a three-cell SQUID} \label{20}

\begin{figure}
	\includegraphics[width=1\columnwidth]{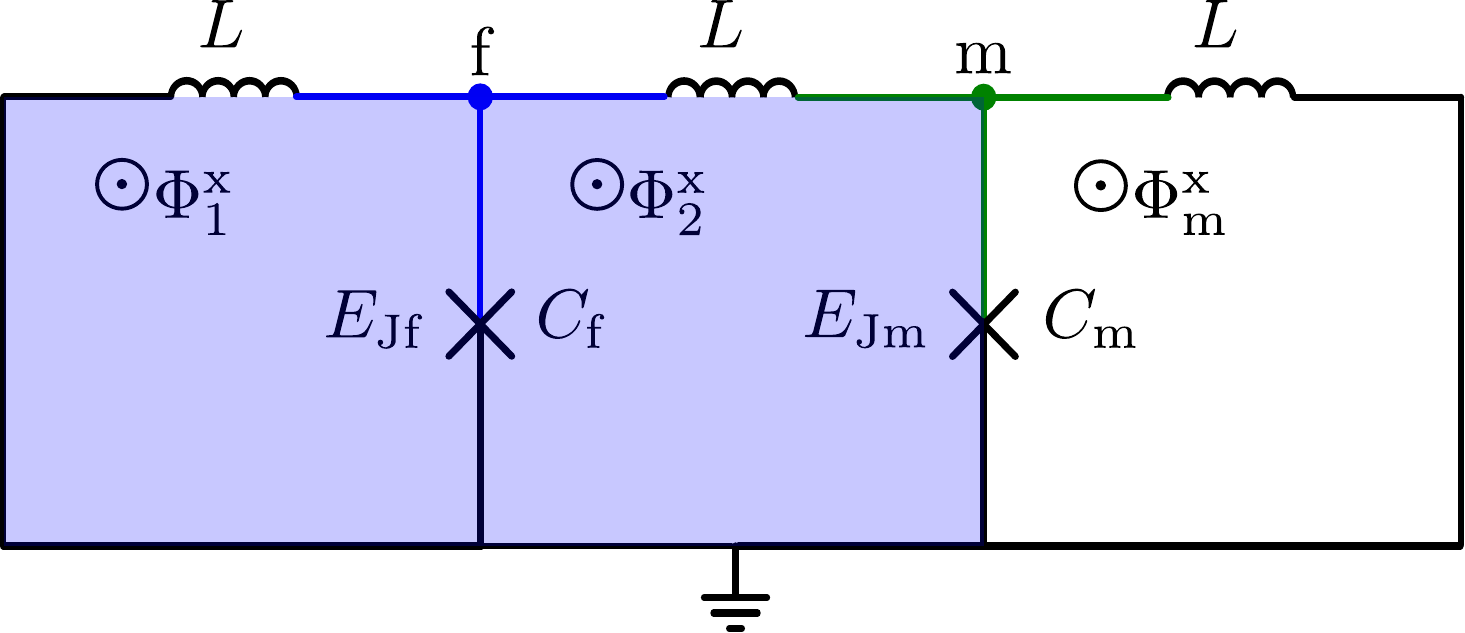}
	\caption{(color online) Equivalent lumped-element circuit for the proposed three-cell SQUID setup: highlighted in different colours are left Josephson junction \textit{f} (blue) and right Josephson junction \textit{m} (green). 
	The inductances, $L$, and externally applied magnetic fields, $\Phi^\textnormal{x}_{1}$, $\Phi^\textnormal{x}_{2}$ and $\Phi^\textnormal{x}_\textnormal{m}$, are shown. }
	\label{fig:full_scematic}
\end{figure}

%\subsection{Full circuit model}
We consider a three-cell SQUID composed of two small Josephson junctions. The schematic of this circuit is shown in Fig.~\ref{fig:full_scematic}. The left (node "\textit{f}" marked with blue lines) and right (node "\textit{m}" marked with green lines) Josephson junctions are characterized by the Josephson coupling energies, $E_{J\textnormal{f}}$ and $E_{J\textnormal{m}}$, accordingly, and large shunt capacitances $C_\textnormal{f}=C_\textnormal{m}=C$. Notice here, that a substitution of a single Josephson junction between the leftmost and rightmost cells by two-junction SQUID allows one to tune the Josephson coupling energies in a wide range.
The Josephson junctions are connected by superconducting loops of high kinetic inductances, $L$ (we consider them to be equal). The externally applied  magnetic fluxes $\Phi^\textnormal{x}_{1}$, $\Phi^\textnormal{x}_{2}$, $\Phi^\textnormal{x}_\textnormal{m}$ penetrate the left, central and right cells of the SQUID, accordingly.   
In particular, this setup allows one to trap a single fluxon in two leftmost (rightmost) SQUID cells, and to use plasma modes excitations in the right (left) Josephson junction for the detection of fluxon location. 

The classical dynamics of Josephson junctions is determined by time-dependent Josephson phases, $\varphi_\textnormal{f}(t)$ and $\varphi_\textnormal{m}(t)$, respectively. 

By making use of the Kirchhoff's current law and magnetic flux quantization in each cell we write the Lagrangian of the three-cell SQUID as follows:
\begin{multline}
L=\frac{1}{2}\frac{C\Phi_0^2}{(2\pi)^2}\dot{\varphi}_\textnormal{f}^2+\frac{1}{2}\frac{C\Phi_0^2}{(2\pi)^2}\dot{\varphi}_\textnormal{m}^2-\\
-E_{J\textnormal{f}}[1-\cos\varphi_\textnormal{f}] -E_{J\textnormal{m}}[1-\cos\varphi_\textnormal{m}]-~~~~~~~~~~~~~~~~~~ \\
-\frac{\Phi_0^2}{2L (2\pi)^2}\left(\varphi_\textnormal{f}+\frac{2\pi}{\Phi_0}\Phi^\textnormal{x}_{1} \right)^2 - \frac{\Phi_0^2}{2L (2\pi)^2}\left(\varphi_\textnormal{m}-\frac{2\pi}{\Phi_0}\Phi^\textnormal{x}_\textnormal{m} \right)^2 -\\
-\frac{\Phi_0^2}{2L (2\pi)^2}\left(\varphi_\textnormal{f}-\varphi_\textnormal{m}-\frac{2\pi}{\Phi_0}\Phi^\textnormal{x}_{2} \right)^2,~~~~~~~~~~~~~~~~~~~~~~~~~
\label{eq2} 
\end{multline} 
where $\Phi_0=h/(2e)$ is the magnetic flux quantum.
Defining the node charges $Q_{\alpha}=\frac{2\pi}{\Phi_0}\partial L/\partial\dot{\varphi}_\alpha$, $(\alpha=\textnormal{f,m})$, we obtain  the circuit Hamiltonian in the following form:
\begin{multline}
\hat{H}=\hat{H}_\textnormal{F}+\hat{H}_\textnormal{M}+\hat{H}_\textnormal{I}, \ \ \ \ \ \ \ \ \ \ \ \ \ \ \ 
\label{eq3}
\end{multline}
\begin{multline}
\hat{H}_\textnormal{F} = \frac{\hat{Q}_\textnormal{f}^2}{2C}+E_{J\textnormal{f}}[1-\cos\hat{\varphi}_\textnormal{f}]+\\ 
+\frac{\Phi_0^2}{L(2\pi)^2} \left (\hat{\varphi}_\textnormal{f}-\pi{\Phi}_{\Delta \textnormal{f}}/\Phi_0 \right)^2 +  \frac{({\Phi}_{\Sigma \textnormal{f}})^2}{2 L},
\label{eq4}
\end{multline}
\begin{multline}
\hat{H}_\textnormal{M} = 
\frac{\hat{Q}_\textnormal{m}^2}{2C}+E_{J\textnormal{m}}[1-\cos\hat{\varphi}_\textnormal{m}]+\\ 
+\frac{\Phi_0^2}{L(2\pi)^2} \left (\hat{\varphi}_\textnormal{m}-\pi{\Phi}_{\Delta \textnormal{m}}/\Phi_0 \right)^2 +  \frac{({\Phi}_{\Sigma \textnormal{m}})^2}{2 L}, 
\label{eq5}
\end{multline}
\begin{multline}
\hat{H}_\textnormal{I} = -\frac{\Phi_0^2}{L (2\pi)^2}\hat{\varphi}_\textnormal{f}\hat{\varphi}_\textnormal{m}.\ \ \ \ \ \ \ \ \ \ \ \ \ \ \
\label{eq6}
\end{multline}

The full Hamiltonian is composed of three parts: $\hat{H}_\textnormal{F}$ ($\hat{H}_\textnormal{M}$)  is the Hamiltonian of autonomous \textit{two}-cell SQUID  with externally applied magnetic fluxes (see the leftmost (rightmost) cells in Fig.~\ref{fig:full_scematic} and Fig.~\ref{fig:two_cells_SQUID}); the Hamiltonian  $\hat{H}_\textnormal{I}$ describes the inductive coupling between the systems $F$ and $M$.
The Hamiltonian $\hat{H}_\textnormal{M}$ can be used to measure the quantum dynamics of the system $F$.
Here we introduce the various combinations of external magnetic fluxes, namely,  ${\Phi}_{\Delta\textnormal{f}}={\Phi}^\textnormal{x}_{2}-{\Phi}^\textnormal{x}_{1}$,  ${\Phi}_{\Sigma\textnormal{f}}={\Phi}^\textnormal{x}_{1}+{\Phi}^\textnormal{x}_{2}$,  ${\Phi}_{\Delta\textnormal{m}}={\Phi}^\textnormal{x}_{m}-{\Phi}^\textnormal{x}_\textnormal{2}$ and  ${\Phi}_{\Sigma\textnormal{m}}={\Phi}^\textnormal{x}_{2}+{\Phi}^\textnormal{x}_\textnormal{m}$.
%of magnetic fluxes in the central and right cells. 
These combinations of magnetic fluxes are not independent but satisfy the condition:  ${\Phi}_{\Sigma \textnormal{m}}-{\Phi}_{\Sigma \textnormal{f}}-{\Phi}_{\Delta \textnormal{f}}-{\Phi}_{\Delta \textnormal{m}}=0$.

\section{QUANTUM DYNAMICS OF A SINGLE FLUXON IN A TWO-CELL SQUID}
\begin{figure}
	\includegraphics[width=1\columnwidth]{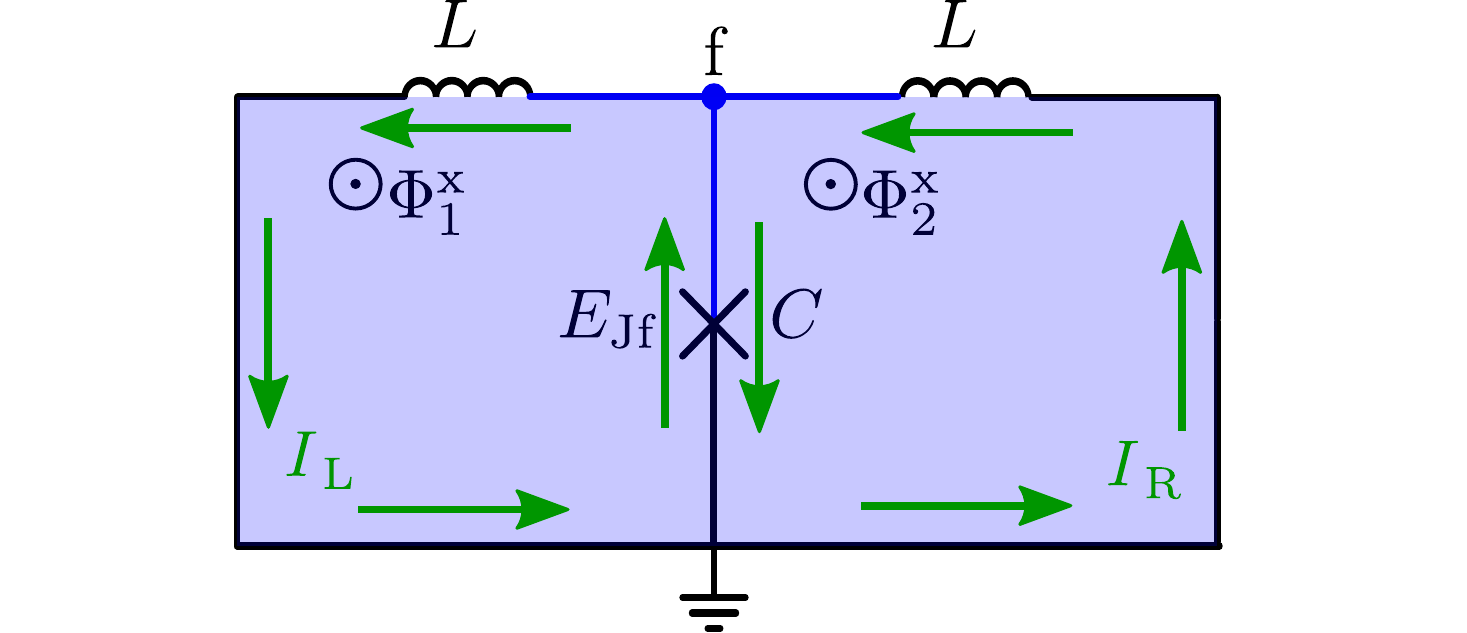}
	\caption{(color online) Circuit model for the two-cell SQUID described by Hamiltonian $\hat{H}_\textnormal{F}$. The left (right) cell is threaded by an external magnetic flux  $\Phi^\textnormal{x}_{1}$ ($\Phi^\textnormal{x}_{2}$). The persistent currents $I_\textnormal{L}$ ($I_\textnormal{R}$) flowing in the left (right) cells are shown. The Josephson coupling energy $E_{J\textnormal{f}}$, shunted capacitance $C$, and the kinetic inductance of superconducting cells $L$ are indicated. }
	\label{fig:two_cells_SQUID}
\end{figure}

%    Figure 1
In this Section, we study the quantum dynamics of a fluxon by analyzing the Hamiltonian (\ref{eq4}). 
Such Hamiltonian determines the quantum dynamics of a two-cell SQUID with a single small Josephson junction. The schematic of such a system is presented in Fig.~\ref{fig:two_cells_SQUID}.
As already mentioned above, high kinetic inductances, $L$ can be established by implementing in the superconducting loops long chains composed of N classical Josephson junctions, forming the so called "superinductance".  With such superinductance the cell inductance $L$ is expressed as $L=N \Phi_0^2/[(2\pi)^2 E_{Ja}]$, where $E_{Ja}$ is a Josephson coupling energy of a single classical Josephson junction in the chain.  In order to suppress the phase fluctuations in the chain we require $E_{Ja}/E_\textnormal{Ca}>>1$, where $E_{Ca}$ is the charging energy of a single Josephson junction in the chain. In this limit the macroscopic quantum tunneling occurs through a small Josephson juction, \textit{f}, only.  The chains of classical Josephson junctions have been widely used as superinductances for fluxonium devices\cite{Manucharyan}. The external flux $\Phi^\textnormal{x}_{1}$ ($\Phi^\textnormal{x}_{2}$) is also applied within the left (right) loop of the SQUID. 

Next, we carried out numerical diagonalization of the Hamiltonian (\ref{eq4}) for various sets of parameters, i.e. varying the values of the Josephson coupling energy $E_{J\textnormal{f}}$, the charging energy $E_{C}=e^2/(2C)$, and the inductive energy $E_L=\Phi^2_0/[L(2\pi)^2]$, to determine the low-lying energy levels and corresponding wave functions (see Appendix A for details). Moreover, by varying externally applied magnetic fluxes $\Phi^\textnormal{x}_{1,2}$ the dependencies of energy levels $E_i$ on $\Phi_\Delta=\Phi^\textnormal{x}_{2}-\Phi^\textnormal{x}_{1}$ were obtained. 

The macroscopic quantum dynamics of a fluxon can be observed in a particular range of parameters only. Firstly, the fluxon "size" has to be reduced. This can be achieved by decreasing of the inductive energy and corresponding increase of the dimensionless parameter $\beta=E_J/E_L \gg 1$.  Secondly, the ratio of Josephson coupling energy $E_{J\textnormal{f}}$ and the charging energy $E_C$ has not to be too large, i.e. $E_{J\textnormal{f}}/E_{C} \geq 1 $. 
Taking in mind these conditions we choose the circuit parameters: $E_C/h=0.5 \ \si{\giga\Hz}$ and $E_L/h=0.15 \ \si{\giga\Hz}$. Note here that these values can be rather easily implemented in experiments with superconducting circuits.

\begin{figure}
	\includegraphics[width=1\columnwidth]{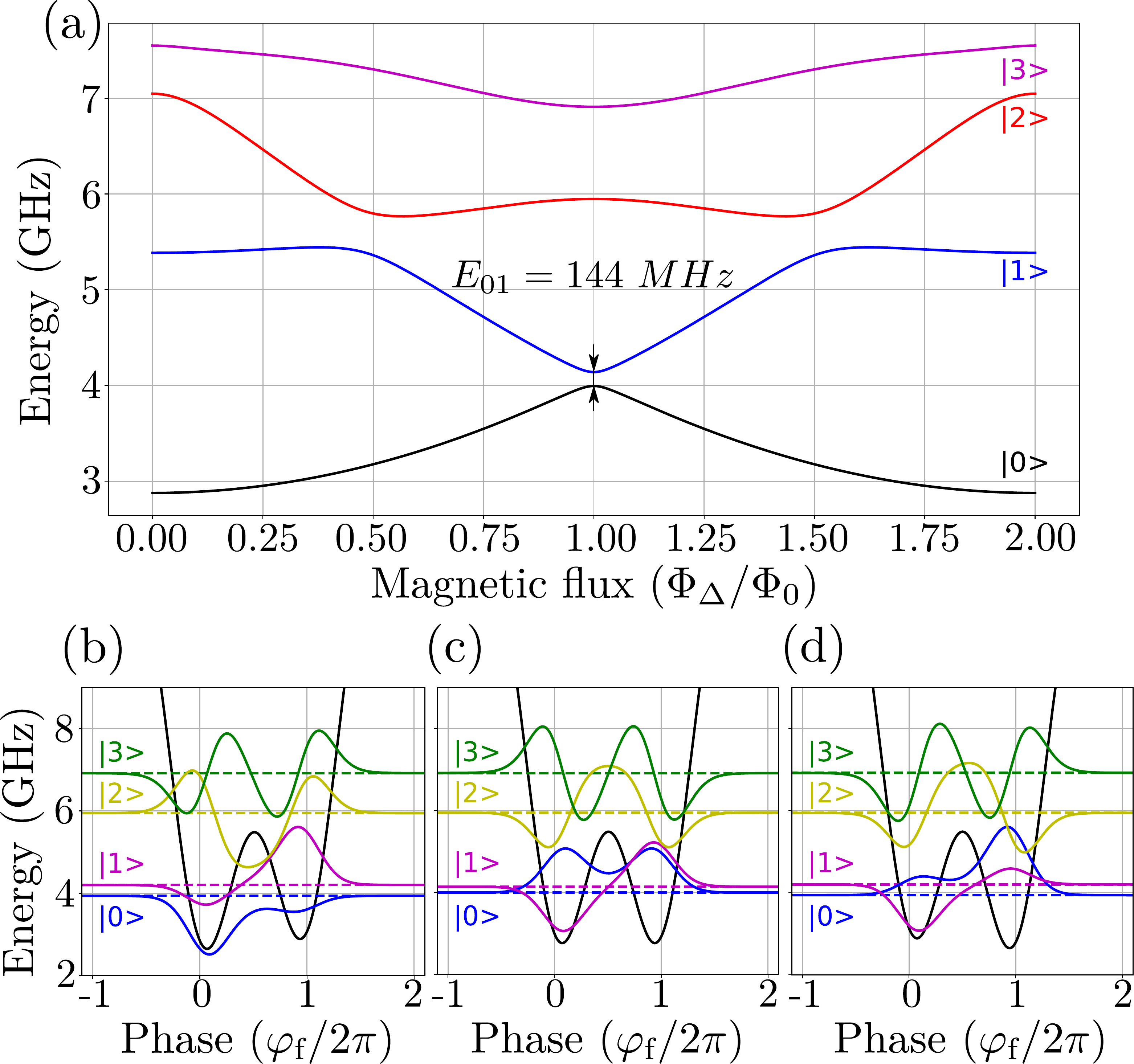}
	\caption{(color online) (a) Low-lying energy levels of the two-cell SQUID vs magnetic flux difference ${\Phi}_{\Delta}$. 
		The parameters $E_{J\textnormal{f}}/h=2 \ \si{\giga\Hz}$, $E_{C}/h=0.5 \ \si{\giga\Hz}$, and $E_L/h=0.15 \ \si{\giga\Hz}$ were used.
		%in the right and left cells for , and ; 
		(b) simulated potential energy landscape/wave functions at ${\Phi}_{\Delta}=0.95 {\Phi}_{0}$; (c) simulated potential energy landscape/wave functions at ${\Phi}_{\Delta}=1.0{\Phi}_{0}$.  (d) simulated potential energy landscape/wave functions at ${\Phi}_{\Delta}=1.05 {\Phi}_{0}$. Here, the states $\ket{0}$ and $\ket{1}$ are nearly degenerate. We put the total magnetic flux in the system ${\Phi}_{\Sigma}=1.0 {\Phi}_{0}$}
	\label{fig:4}
\end{figure}

Here, we first present the computed energy spectrum  for $E_{J\textnormal{f}}/h=2 \ \si{\giga\Hz}$ ($\beta \simeq 13.3$) as a function of magnetic flux difference ${\Phi}_{\Delta}$. Such dependencies $E_i(\Phi_\Delta)$  are plotted in Fig.~\ref{fig:4}a for the lowest four energy levels of the system. As one can see, the energy states $\ket{0}$ and $\ket{1}$ become nearly degenerate as $\Phi_\Delta=1.0 \Phi_0$, and the splitting $\Delta E_{01}(\Phi_\Delta)=E_1-E_0$ increases as  the magnetic flux difference $\Phi_\Delta$ deviates from the degeneracy point, $\Phi_\Delta=1.0 \Phi_0$.
In Fig.~\ref{fig:4}(b,c,d) we show low-lying energy levels accurately positioned in the potential profile ${U}(\varphi_\textnormal{f})$, along with their calculated wave functions at $0.95 {\Phi}_{0}$, $1.0 {\Phi}_{0}$ and $1.05 {\Phi}_{0}$ magnetic flux difference values. 

\begin{figure}
	\includegraphics[width=1\columnwidth]{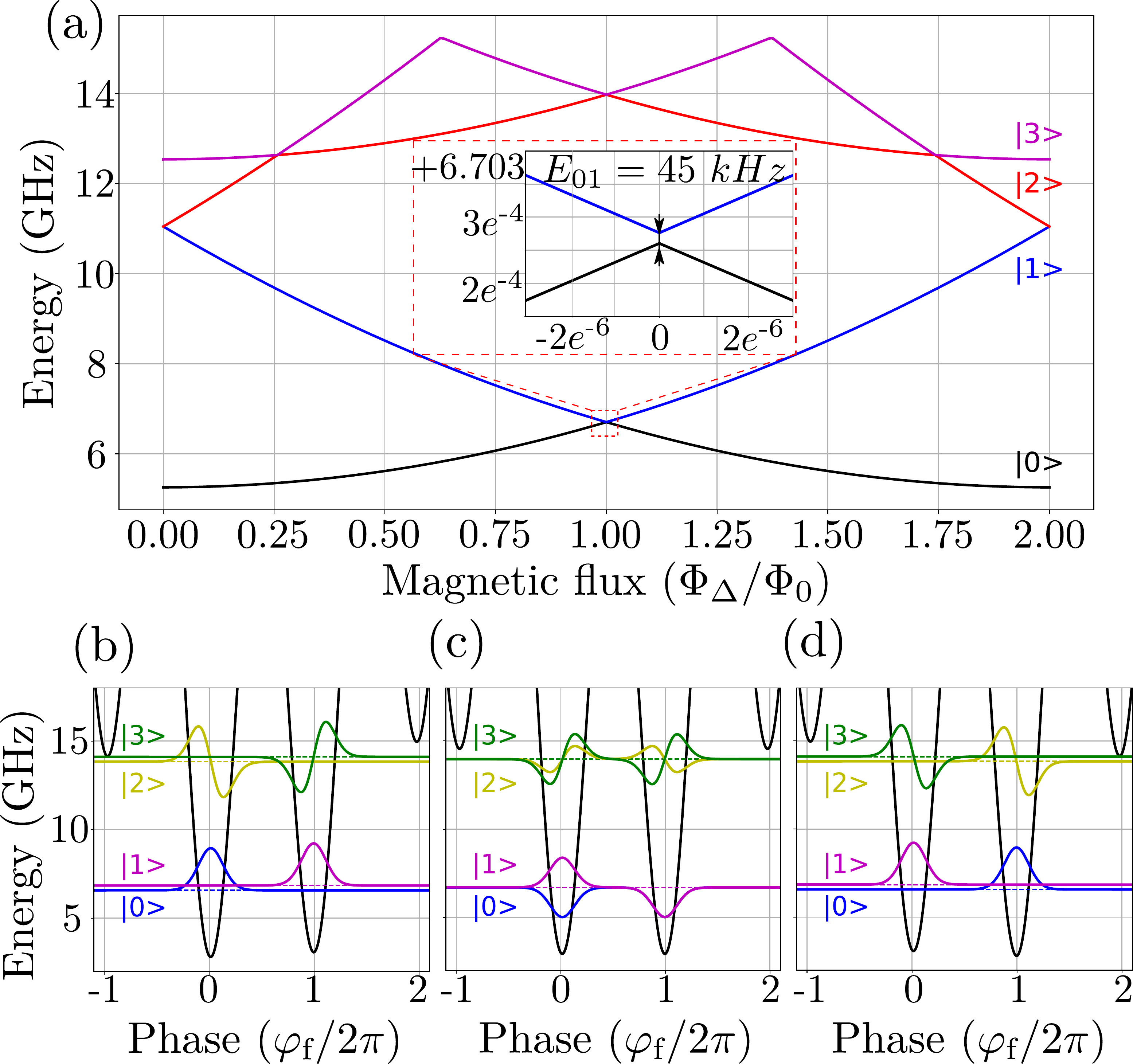}
	\caption{(color online) (a) Low-lying energy levels of the two-cell SQUID vs magnetic flux difference ${\Phi}_{\Delta}$. 
		The parameters $E_{J\textnormal{f}}/h=15 \ \si{\giga\Hz}$, $E_{C}/h=0.5 \ \si{\giga\Hz}$, and $E_L/h=0.15 \ \si{\giga\Hz}$ were used.
		%in the right and left cells for , and ; 
		(b) simulated potential energy landscape/wave functions at ${\Phi}_{\Delta}=0.95 {\Phi}_{0}$; (c) simulated potential energy landscape/wave functions at ${\Phi}_{\Delta}=1.0 {\Phi}_{0}$. (d) simulated potential energy landscape/wave functions at ${\Phi}_{\Delta}=1.05 {\Phi}_{0}$. Here, the states $\ket{0}$ and $\ket{1}$ are nearly degenerate. We put the total magnetic flux in the system ${\Phi}_{\Sigma}=1.0 {\Phi}_{0}$}
	\label{fig:5}
\end{figure}

\begin{figure}
	\includegraphics[width=1\columnwidth]{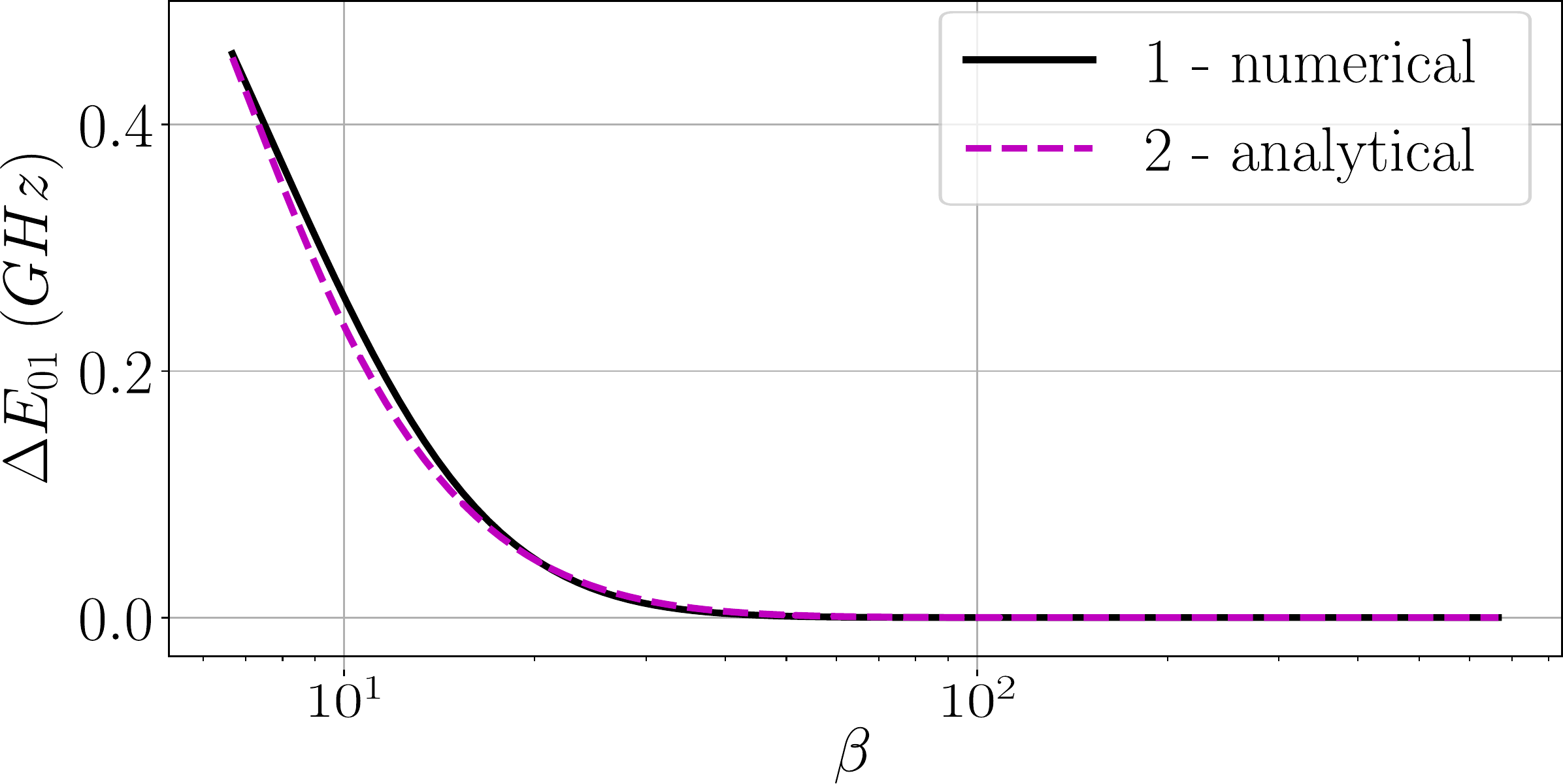}
	\caption{(color online) The dependence of energy level splitting $\Delta=\Delta E_{01}(\Phi_\Delta=1.0 \Phi_0)/h$  between qubit states $\ket{0}$ and $\ket{1}$ on the parameter $\beta$: solid  line - numerical calculations; dashed line-analytical approach (see. Appendix \ref{200}). The parameters were chosen as $E_C/h=1 \ \si{\giga\Hz}$, $E_L/h=0.15 \ \si{\giga\Hz}$ and $E_{J\textnormal{f}}/h=1\div100 \ \si{\giga\Hz}$.} 
	\label{fig:6}
\end{figure}

An increase of the Josephson coupling energy $E_{J\textnormal{f}}$ leads to a strong decrease of the energy level separation, $\Delta$, at the degeneracy point, $\Phi_\Delta=1.0 \Phi_0$, i.e. $\Delta=\Delta E_{01}(\Phi_\Delta=1.0 \Phi_0)$. The dependence of low-lying energy levels on the magnetic flux, $E_i (\Phi_\Delta)$, for $E_{J\textnormal{f}}/h=15 \ \si{\giga\Hz}$ ($\beta=100$) is presented in Fig.~\ref{fig:5}(a), and in Fig.~\ref{fig:5}(b,c,d) we show low-lying energy levels, the potential profile ${U}({\phi}_\textnormal{f})$, along with their calculated wave functions. The numerically calculated dependence of energy level splitting $\Delta$ on the parameter $\beta$ is shown in Fig.~\ref{fig:6} by solid line.

\begin{figure}
	\includegraphics[width=1\columnwidth]{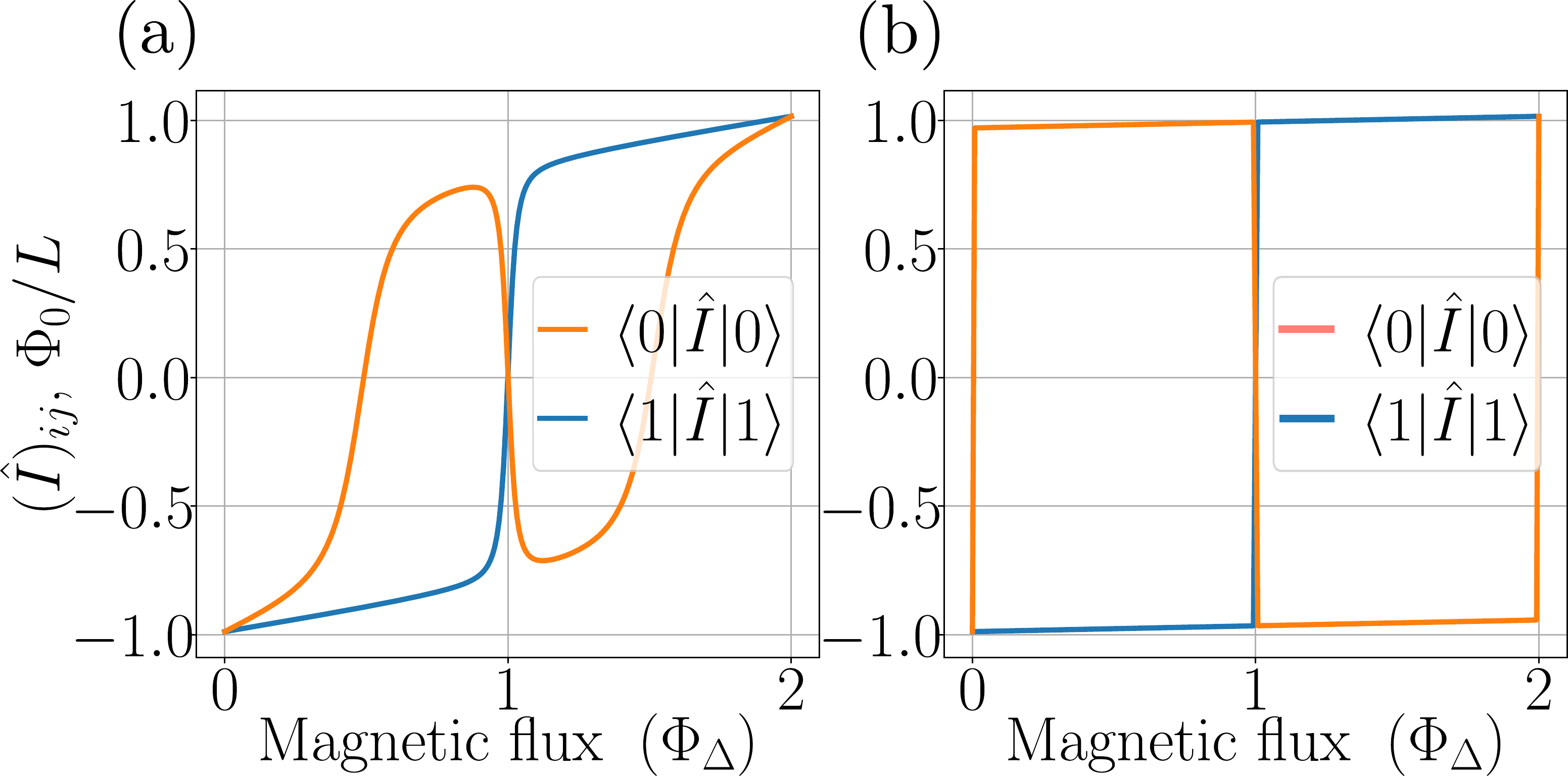}
	\caption{(color online) The dependence of the matrix elements of current operator, i.e.  $\hat{I}_{00}$ (orange lines) and  $\hat{I}_{11}$ (blue line) on the magnetic flux difference $\Phi_{\Delta}$ for different Josephson coupling energies: (a) $E_{J\textnormal{f}}/h=2 \ \si{\giga\Hz}$; (b) $E_{J\textnormal{f}}/h=15 \ \si{\giga\Hz}$. Other parameters were chosen as  $E_{C}/h=0.5 \ \si{\giga\Hz}$ and $E_L/h=0.15 \ \si{\giga\Hz}$.}
	\label{fig:7}
\end{figure}

In most interesting region of magnetic fluxes $\Phi_\Delta \simeq 1.0 \Phi_0$ and for large parameter $\beta \gg 1$ the 
Hamiltonian (\ref{eq4}) can be linearized, and it is well approximated as
\begin{multline}
\hat{H}_{\textnormal{F}}= \hat{H}_\textnormal{F}(\Phi_{0}) + \left.\delta \Phi_{\Delta} \frac{\partial{\hat{H}_\textnormal{F}(\Phi_{\Delta})}}{\partial{\Phi_{\Delta}}} \right|_{ \Phi_{0}} =\\
=\hat{H}_\textnormal{F}(\Phi_{0}) +\delta \Phi_{\Delta} \hat{I},
\label{eq10}
\end{multline}
where $\delta \Phi_{\Delta}=(\Phi_{\Delta} - \Phi_{0})$ and 
the operator of current $I$ flowing through the Josephson junction is determined as $\hat{I}=\frac{\Phi_0}{2\pi L}(\pi-\hat{\varphi}_\textnormal{f})$. By making use of the Kirchhoff's law the current $I$ can be expressed as $I=I_\textnormal{L}-I_\textnormal{R}$, where the currents $I_L$ ($I_R$) flowing through the left (right) superconducting cells.
%are $I_L=\frac{(\vatphi_f-\pi{\Phi}^x_{1})}{2L}$ and right %$\hat{I}_\textnormal{R}=\frac{(\hat{\Phi}_\textnormal{f}-{\Phi}^\tex%tnormal{x}_{1})}{2L_1}$ inductances.
The dependence of matrix elements 
%of the current operator 
$\hat{I}_{00}=\bra{0}\hat{I}\ket{0}$ and $\hat{I}_{11}=\bra{1} \hat{I} \ket{1}$ for the eigenstates $\ket{0}$ and $\ket{1}$ on the magnetic flux difference $\Phi_{\Delta}$ was numerically calculated, and it is presented in 
Fig.~\ref{fig:7} for two different parameters of $E_{J\textnormal{f}}$. Near the degeneracy point ($\Phi_{\Delta}=\Phi_{0}$) the matrix elements have opposite signs, i.e. $\hat{I}_{00}=-\hat{I}_{11}$, and for extremely large values of $\beta$  the matrix elements $|\hat{I}_{ii}| \approx \pm \frac{\Phi_0}{L}, i=0,1$ (see, Fig.~\ref{fig:7}b).

Thus, from the presented  analysis one can conclude that large positive (negative) values of the current $I$ correspond  to persistent superconducting currents flowing in the left (right) cells. These persistent currents determine the location of a trapped fluxon in the left (right) cell. Therefore, the quantum dynamics of  two low-lying eigenstates can be  described in two-state fluxon location basis, $|L>$ and $|R>$. In this basis the Hamiltonian of a single fluxon has a well-known single qubit form:
\begin{equation}
\hat{H}_{\textnormal{2cells-MF}}= \frac{\Delta}{2}\hat{\sigma}^\textnormal{x}+\frac{\epsilon}{2}\hat{\sigma}^\textnormal{z},
\label{HamiltonianMF}
\end{equation}
where $\epsilon_0=\delta \Phi_{\Delta} \frac{\Phi_0}{2 L}$. This Hamiltonian determines the energy levels splitting as 
$\Delta E_{01}=\sqrt{\Delta^2+\epsilon^2}$. 

As the magnetic flux difference $\Phi_\Delta$ is tuned to the degeneracy point, the matrix elements of the current operator $\hat{I}_{ii}|, i=0,1$ goes to zero value, and it indicates that the dynamics of a single quantum fluxon is determined by the quantum tunneling between neighboring cells, and the coherent \textit{quantum beats} (coherent oscillations of the probability amplitude) with the frequency $f_{qb}=\Delta{E}_{01}/h$ occur. 
%The inter-well transitions in such systems are called fluxons. 

The dynamics of a single fluxon in a two-cell SQUID can be also understood by analyzing the classical potential profile $U(\varphi_\textnormal{f})$. In the regime of $\beta \gg 1$ the potential  ${U}(\varphi_\textnormal{f})$ consists of Josephson wells, whose depth and elevation depend   
on $E_{J\textnormal{f}}$ (${E}_L$ is fixed) and ${\Phi}_{\Delta}$, respectively (see, Figs.~\ref{fig:4}b,c,d and \ref{fig:5}b,c,d). In classical regime, the Josephson phase $\varphi_\textnormal{f}$ is located at minimums of $U(\varphi_\textnormal{f})$. Thus, the values of  $\varphi_\textnormal{f} \simeq 0$ ($\varphi_\textnormal{f} \simeq 2\pi$) correspond to the localization of fluxon in left (right) cells. In the quantum regime the non-zero value of charging energy results in the coherent tunneling and quantum beats of trapped fluxon 
between neighboring cells. In the limit of $\beta \gg 1$, by making use of the quasi-classical approximation the dependence of quantum beats frequency $f_{qb}(\Phi_\Delta=1.0 \Phi_0)$ on the inductance parameter $\beta$ was calculated analytically (see details in Appendix \ref{200}) . Such dependence is presented in Fig.~\ref{fig:7} by dashed line.

\section{Quantum dynamics of a single FLUXON in a three-cell SQUID}\label{30}
Next, we analyze the quantum dynamics of a single trapped fluxon in a three-cell SQUID of high kinetic inductance (see, Fig.~\ref{fig:full_scematic}). The coherent quantum dynamics of such a system is determined by the Hamiltonian (\ref{eq3})-(\ref{eq6}), and the trapped fluxon displays two quantum-mechanical effects, namely, the coherent quantum tunneling of the fluxon between adjacent cells, and the excitations of plasma  oscillations in Josephson junctions, \textit{f} and \textit{m}. As we discussed in Section III the coherent quantum tunneling of fluxons can be enhanced by manipulation of  the Josephson coupling energies as  $E_{J\textnormal{f}}, E_{J\textnormal{m}} \geq E_C$ (see, the Fig.~\ref{fig:4}). In this Section we consider an opposite case, i.e.  $E_{J\textnormal{f}} \gg  E_C$, and  $E_{J\textnormal{m}} \gg  E_C$, as if the coherent tunneling of fluxons is strongly suppressed, and the fluxon is localized in a one cell of the SQUID. 

\subsection{Low-lying energy spectrum and reduced Hamiltonian: plasma modes excitations}
To analyze quantitatively the low-lying spectrum of a strongly localized fluxon, we carry out the numerical calculation (see details in Appendix A) of the low-lying eigenvalues and eigenfunctions of the Hamiltonian (\ref{eq3}) for a particular set of parameters: $\Phi^\textnormal{x}_\textnormal{m}=0$, $\Phi_{\Sigma\textnormal{f}}=\Phi^\textnormal{x}_\textnormal{1}+\Phi^\textnormal{x}_\textnormal{2}=0$. The magnetic flux difference $\Phi_{\Delta\textnormal{f}}=\Phi^\textnormal{x}_\textnormal{2}-\Phi^\textnormal{x}_\textnormal{1}$ was varied in the calculations. In order to suppress the quantum tunneling and strongly localize a single fluxon in a one of three cells, the Josephson coupling energies, $E_{J\textnormal{f}}/h= 20 \  \si{\giga\Hz}$, $E_{J\textnormal{m}}/h= 22 \ \si{\giga\Hz}$ were chosen to be much larger than the charging energy, $E_{\textnormal{C}}/h=0.5 \ \si{\giga\Hz}$. Here, we assume that the Josephson coupling energies of Josephson junctions \textit{f} and \textit{m} are slightly different to remove the degeneracy of excited levels.

\begin{figure}
	\includegraphics[width=1\columnwidth]{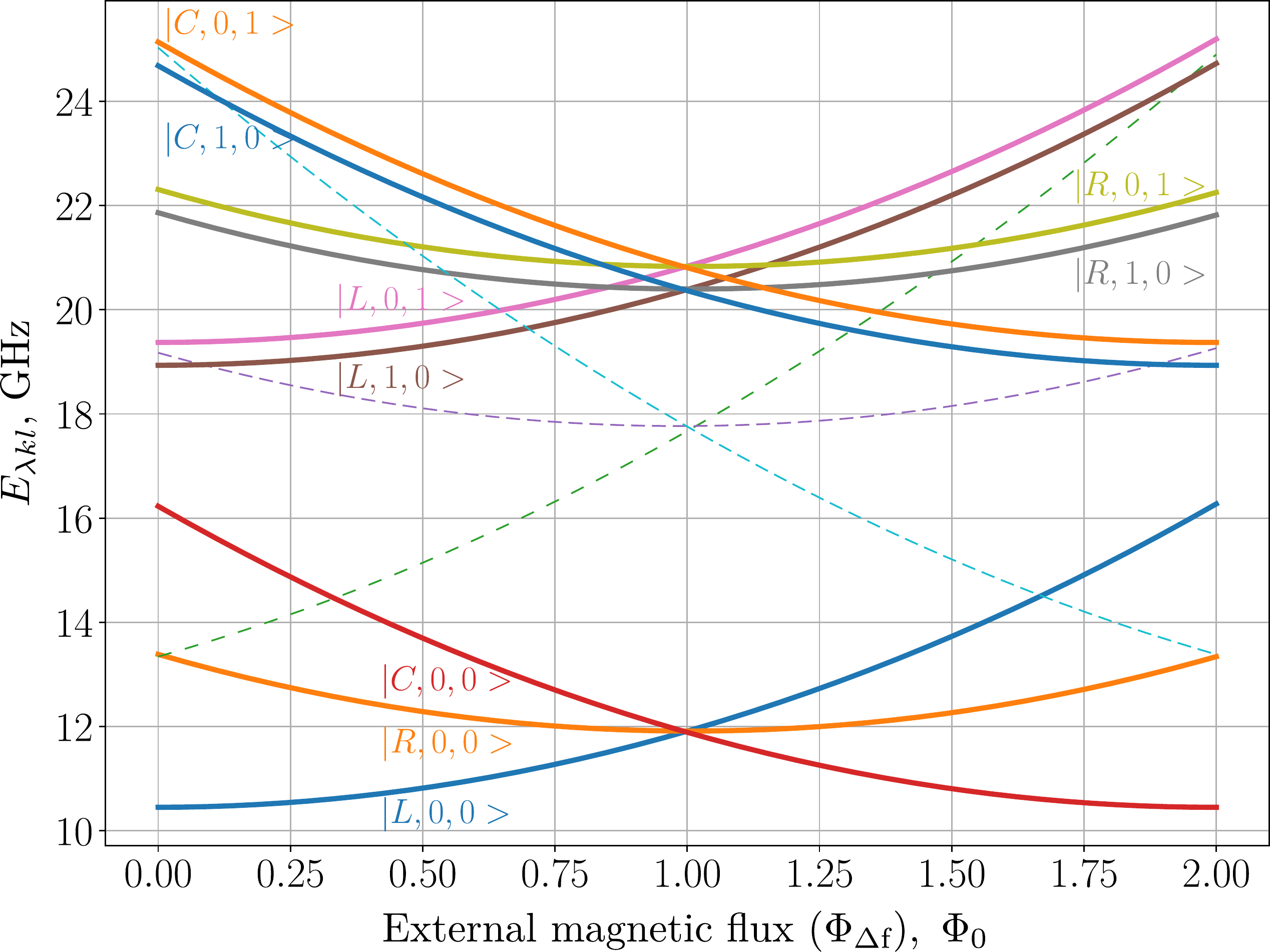}
	\caption{(color online) The dependence of the energy spectrum  $E_{\lambda k,\ell}$ of a single fluxon trapped in a three-cell SQUID  on the magnetic flux difference ${\Phi}_{\Delta\textnormal{f}}$. The signatures $\ket{\lambda,k,\ell}$ indicate the corresponding eigenstates. The parameters of a system were chosen as $E_{J\textnormal{f}}/h=20 \ \si{\giga\Hz}$, $E_{J\textnormal{m}}/h=22 \ \si{\giga\Hz}$, $E_{C}/h=0.5 \ \si{\giga\Hz}$, and  $E_L/h=0.15 \ \si{\giga\Hz}$. Dashed lines show energy levels for the case when two fluxons and one anti-fluxon are present in the system.}
	\label{fig:Full_spectr}
\end{figure}

The low-lying eigenvalues and corresponding eigenstates of the Hamiltonian $\hat{H}$ (\ref{eq3}) on a one-fluxon manifold can be conveniently described in the eigenstates basis labeled as $\ket{\lambda, k, \ell}=\ket{\lambda} \otimes \ket{ k}_\textnormal{f} \otimes \ket{\ell}_\textnormal{m}$, where $\lambda$ represents the fluxon location in the left ($L$), central ($C$) or right ($R$) cell; $k$ and $\ell$ are the plasma mode occupancy associated with variables $\varphi_\textnormal{f}$ and $\varphi_\textnormal{m}$, accordingly. Here, we only consider the low-lying excitations of plasma modes as the values of $k,\ell=0,1$.  
The numerically calculated dependence of low-lying energy spectrum on the magnetic flux difference $\Phi_{\Delta\textnormal{f}}$ is presented in Fig.~\ref{fig:Full_spectr}, and one can see that near the degeneracy point (${\Phi}_{\Delta\textnormal{f}}={\Phi}_0$) the energy levels form three distinguished groups: the first group of eigenstates (the energy close to $12.0 \ \si{\giga\Hz}$), $\ket{L,0,0}$, $\ket{C,0,0}$, $\ket{R,0,0}$, corresponds to absence of plasma excitations  in both Josephson junctions;  the second group (energy close to $20.4 \ \si{\giga\Hz}$), $\ket{L,1,0}$, $\ket{C,1,0}$, $\ket{R,1,0}$ corresponds to a single plasma mode excitation in the Josephson junction \textit{f};  the third group (energy close to $20.9 \ \si{\giga\Hz}$), $\ket{L,0,1}$, $\ket{C,0,1}$, $\ket{R,0,1}$ corresponds to a single plasma mode excitation in the Josephson junction \textit{m}. 
 
\begin{figure}
	\includegraphics[width=0.85\columnwidth]{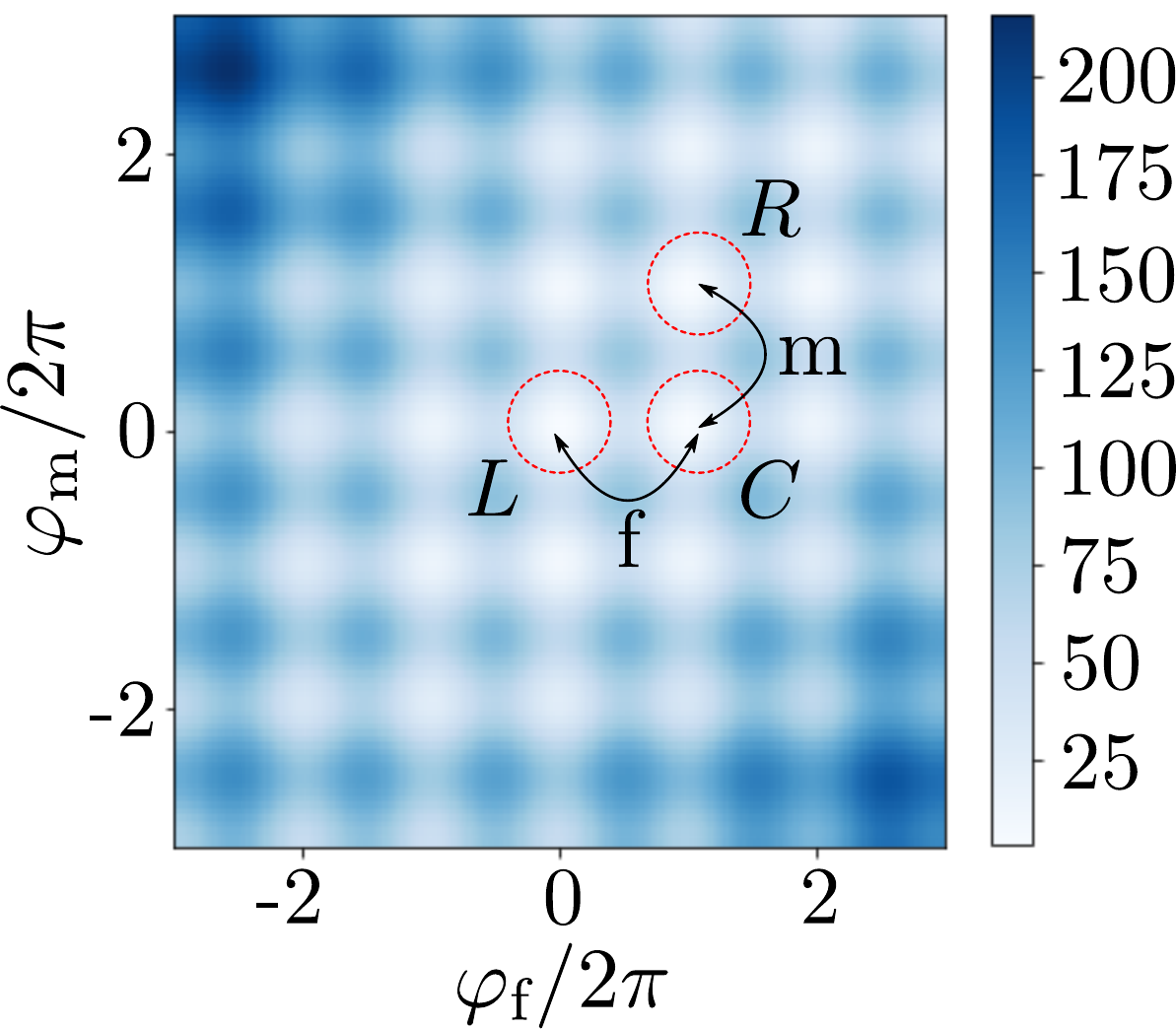}
	\caption{(color online) Two-dimensional landscape of the potential energy $U(\varphi_\textnormal{f}, \varphi_\textnormal{m})$ for a three-cell SQUID on the degeneracy point ${\Phi}_{\Delta\textnormal{f}}={\Phi}_0$. The red circles indicate three lowest potential wells. Signs $L, C, R$ indicate the localization of the fluxon in the corresponding cell  of the circuit. The parameters were chosen as $E_{J\textnormal{f}}/h=20 \ \si{\giga\Hz}$, $E_{J\textnormal{m}}/h=22 \ \si{\giga\Hz}$, $E_L/h=0.15 \ \si{\giga\Hz}$. }
	\label{fig:Potential}
\end{figure}

The quantum dynamics of a single fluxon trapped in the three-cell SQUID can be qualitatively understood by consideration of a two-dimensional landscape of the potential energy $U(\varphi_\textnormal{f}, \varphi_\textnormal{m})$ demonstrating a multi-well shape. 
For the magnetic flux difference $\Phi_{\Delta\textnormal{f}}=1.0 \Phi_0$ the potential energy landscape $U(\varphi_\textnormal{f}, \varphi_\textnormal{f})$ is presented in Fig.~\ref{fig:Potential}. The three lowest potential wells are indicated by 
%dashed 
red circles and denoted as $L, C, R$ because of the states located close to the minima of the potential energy correspond to the localization of a single fluxon in a particular cell. The arrows indicate fluxon tunneling between adjacent cells through Josephson junctions \textit{f} and \textit{m}.
 
\begin{figure}
	\includegraphics[width=1\columnwidth]{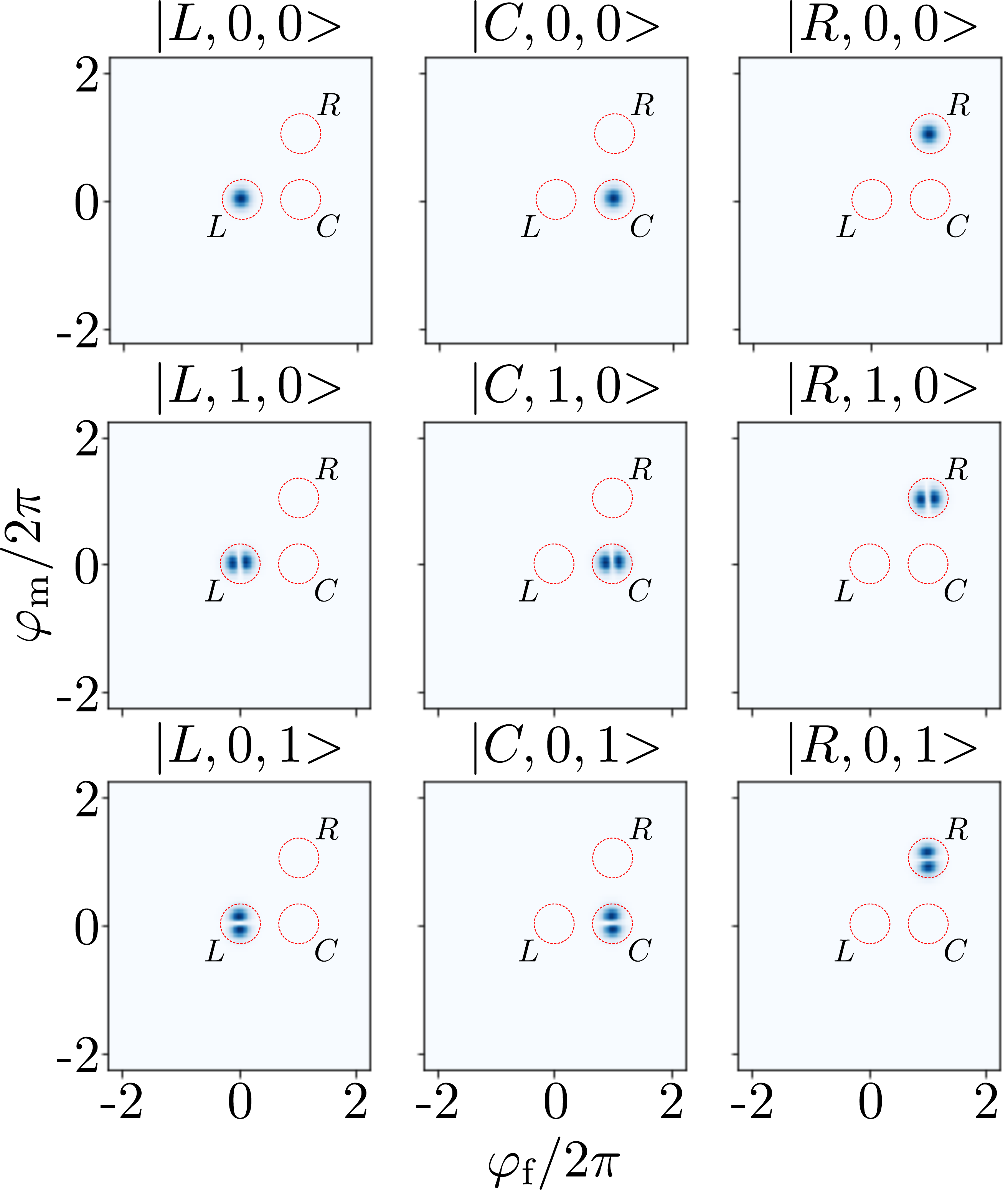}
	\caption{(color online) Two-dimensional landscape of the probability density of eigenstates for the three groups 
		of states. The magnetic flux difference   ${\Phi}_{\Delta\textnormal{f}}=1.0 {\Phi}_0$. Red circles also show three lowest potential wells  (same as in Fig.~\ref{fig:Potential}). Signatures $\ket{\lambda, k, l}$ indicate the corresponding state as in Fig.~\ref{fig:Full_spectr}.}
	\label{fig:waves}
\end{figure}

In Fig.~\ref{fig:waves} we plot the two-dimensional landscape of probability of density for the eigenstates corresponding to three energy levels groups. The red circles on plots also indicate the locations of the fluxon as in Fig.~\ref{fig:Potential}. The first row in the Fig.~\ref{fig:waves} displays the eigenstates in the absence of plasma excitations, while the second (third) row displays the eigenstates with a single plasma excitation on the Josephson junction \textit{f} (\textit{m}). 
Thus, one can describe the general eigenstate $\ket{\lambda, k,\ell}$ of a single trapped fluxon in a three-cell SQUID as a \textit{qutrit} with the basis states $\ket{L}$, $\ket{C}$ and $\ket{R}$, interacting with \textit{two transmon qubits } \textit{f} and \textit{m}. 

Analyzing the energy spectrum presented in Fig.~\ref{fig:Full_spectr} we obtain a peculiar effect. At the degeneracy point the various transitions associated with the excitation of plasma modes on \textit{f}-Josephson junction demonstrate the following frequencies:
 $E_{\ket{L,0,0}\,\to\,\ket{L,1,0}}=E_{\ket{C,0,0}\,\to\,\ket{C,1,0}}\approx8.4771 \ \si{\giga\Hz}$, but  $E_{\ket{R,0,0}\,\to\,\ket{R,1,0}}\approx8.4844 \ \si{\giga\Hz}$.  It indicates that the plasma modes excited on \textit{f}-Josephson junction have the same frequencies for the fluxon localized in left and central cells, but the frequency of plasma mode for fluxon localized in the right cell is slightly ($7.3 \ \si{\mega\Hz}$) higher. 
Similarly the transitions associated with the excitation of plasma modes on \textit{m}-Josephson junction demonstrate the following frequencies: $E_{\ket{C,0,0}\,\to\,\ket{C,0,1}}=E_{\ket{R,0,0}\,\to\,\ket{R,0,1}}\approx8.9181 \ \si{\giga\Hz}$ but $E_{\ket{L,0,0}\,\to\,\ket{L,0,1}}\approx8.9242 \ \si{\giga\Hz}$. Here the plasma modes excited on \textit{m}-Josephson junction have the same frequencies for fluxon localized in central and right cells but the frequency of plasma mode for fluxon localized in the left cell is slightly ($6.1 \ \si{\mega\Hz}$) higher.

Now, we can describe the one-fluxon manifold with reduced Hamiltonian  of two transmon qubits coupled with a qutrit system:
\begin{multline}
	\hat{H}_{Qb-Qtr}= \frac{1}{2}\hbar \omega_{p\textnormal{f}} \hat{\sigma}^\textnormal{z}_\textnormal{f} + \frac{1}{2}\hbar \omega_{p\textnormal{m}} \hat{\sigma}^\textnormal{z}_\textnormal{m} +\\
	\epsilon_L\ket{L}\bra{L}+\epsilon_C\ket{C}\bra{C}+\epsilon_R\ket{R}\bra{R}+\\
	+J^{\textnormal{z}}_{\textnormal{f}}\hat{\sigma}^\textnormal{z}_\textnormal{f}\ket{R}\bra{R} +J^{\textnormal{z}}_\textnormal{m}\hat{\sigma}^\textnormal{z}_\textnormal{m}\ket{L}\bra{L},
	\label{eq19}
\end{multline}
where $\hbar \omega_{p\textnormal{f(m)}} \simeq \sqrt{8E_{J\textnormal{f(m)}}E_C}$ are the Josephson plasma frequencies of the \textit{f(m)}-Josephson junctions, $\hat{\sigma}^\textnormal{x,y,z}_{\alpha}$ are corresponding Pauli matrices for \textit{f}- ($\alpha=\textnormal{f}$) and \textit{m}- ($\alpha=\textnormal{m}$) qubits. The states of the qutrit are denoted by $\lambda$ ($\lambda=L,C,R$), and the parameters $\epsilon_L$, $\epsilon_C$ and $\epsilon_R$  are magnetic flux dependent energies of states $\ket{L}$, $\ket{C}$ and $\ket{R}$, accordingly.

The parameters of dispersive interactions $J^{\textnormal{z}}_{\textnormal{f(m)}}$ are obtained as follows. The interaction  between the transmon qubits and the qutrit is determined by the interaction Hamiltonian, $\hat H_I$, (see, Eq. (\ref{eq6})), and such interaction contains of  off-diagonal terms in the eigenbasis of transmon-qubits, i.e. 
$g_{\alpha} \hat{\sigma}^{\textnormal{x}}_\alpha$. The interaction strengths $g_\textnormal{f(m)}$  are obtained as
$g_{\textnormal{f(m)}} \simeq 2\pi E_L \sqrt{\hbar \omega_{p\textnormal{f(m)}}/[2E_{J\textnormal{f(m)}}]} \ll \hbar \omega_{p\textnormal{f(m)}}$. Using the perturbation approach we derive the Hamiltonian (\ref{eq19}) where the dispersive interaction has a diagonal form in the eigenbasis of transmon-qubits, and the interaction strengths are determined as  
$J^{\textnormal{z}}_{\textnormal{f(m)}} = 2g^2_{\textnormal{f(m)}}/[\hbar \omega_{p\textnormal{f(m)}}]$ and they differ on small values (as shown above $J^{\textnormal{z}}_{\textnormal{f}}/h=7.3 \ \si{\mega\Hz}$ and $J^{\textnormal{z}}_{\textnormal{m}}/h=6.1 \ \si{\mega\Hz}$).

From this Hamiltonian one can conclude that the positions of fluxon in the cells adjacent to the measured node couldn't be distinguished via monitoring plasma mode transitions. However, the location of fluxon in a distant cell lead to a small frequency shift for this particular measured transition. That allows us to apply the traditional non-demolition measurement  technique \cite{Nondemolution,Nondemolution2} of transmon qubits for distinguish fluxon states in different cells. In other words, the fluxon located in leftmost cell will shift the mode of junction \textit{m}, while the fluxon sitting in the rightmost cell will shift the mode of junction \textit{f}.

\subsection{Readout of fluxon states}
To perform time resolved measurements of the quantum beats of a single fluxon trapped in the three-cell SQUID one can use the following procedure consisting of the three steps: a) initialization of a system, i.e. the trapping of a single fluxon in the particular cell; b) inducing the quantum beats of a single trapped fluxon between \textit{two} cells. It can be achieved by suppression of the effective Josephson energy of a single Josephson junction, i.e. $E_{J\textnormal{f}} \geq E_C$ or $E_{J\textnormal{m}} \geq E_C$; c) stopping quantum beats by strong fluxon localization. It can be achieved by increasing the Josephson energies of both Josephson junctions  ($E_{J\textnormal{f(m)}} \gg E_C$). The rapid change of Josephson energies in a wide region is realized by the standard method, i.e. substitution of a single Josephson junction by a two-junction SQUID and variation of the magnetic field penetrating such SQUID.
Control of the fluxon's position is possible by changing the magnetic flux difference in neighboring SQUID cells, as can be seen from the comparison of Fig.\ref{fig:5}b ($\Phi_{\Delta}=0.95\Phi_0$ and fluxon localized in left cell) and Fig.\ref{fig:5}d ($\Phi_{\Delta}=1.05\Phi_0$ and fluxon localized in right cell).    

Once the above manipulations are completed, the dispersive readout of the position of fluxon in a single cell can be carried out. The dispersive 
interactions $J^{\textnormal{z}}_{\textnormal{f(m)}}$ lead to an additional phase accumulation (a single plasmon excitation) of the corresponding transmon qubit as the fluxon is located in the right or left cell, respectively. 

\begin{figure}
	\includegraphics[width=1\columnwidth]{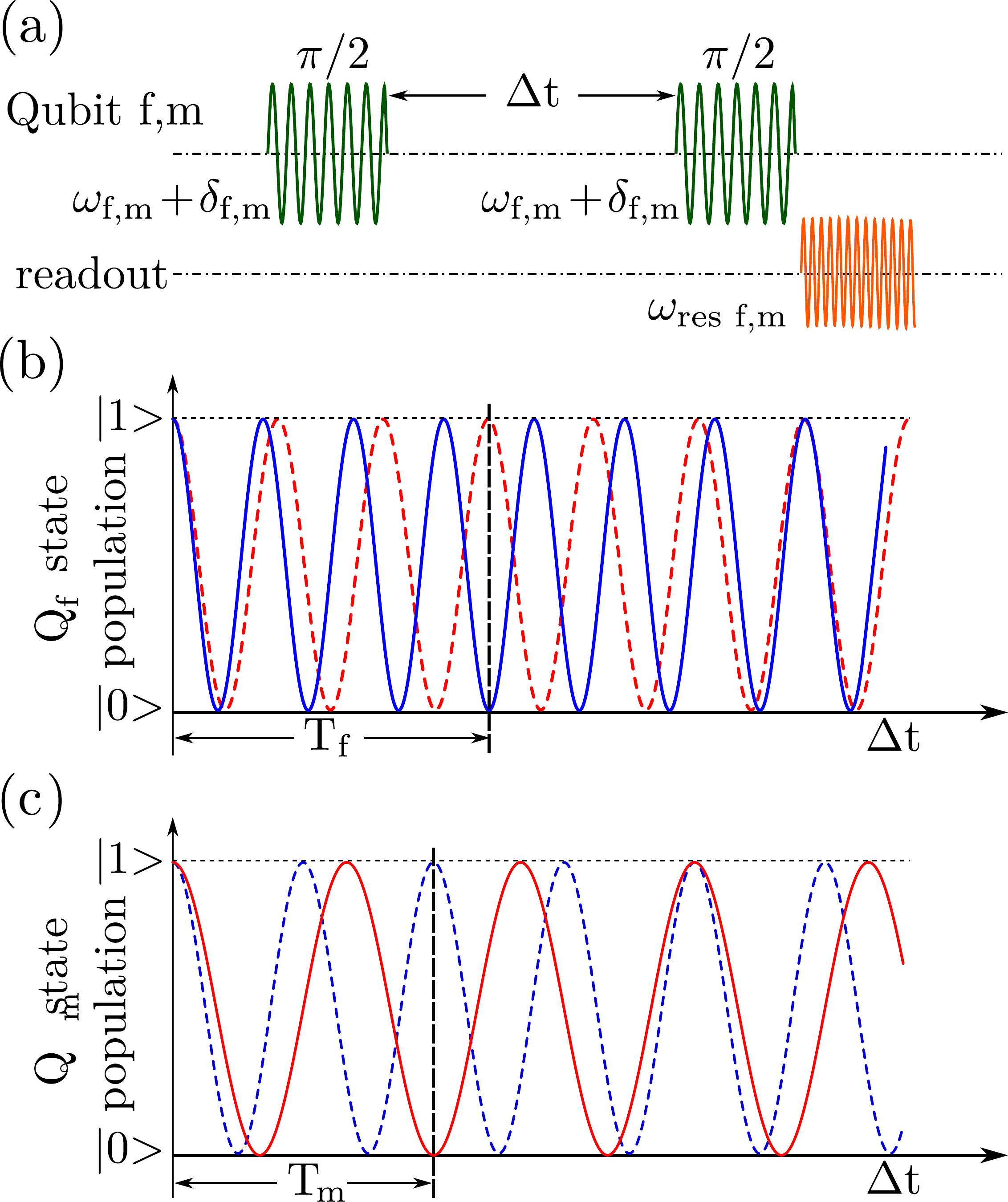}
	\caption{(color online) (a) - The Ramsey pulse sequence used to calibrate the CPS; (b),(c) - Examples of the Ramsey fringes for the transmon-qubit \textit{f} (the transmon-qubit \textit{m})	(not in the scale). The Ramsey fringes corresponding to the fluxon position in the left (right) cell, are indicated by blue (red) line.
		%state population during the calibration 
		%we can see different Ramsey fringes for starting with blue curves) versus %fluxon localized in right cell (red curves). 
		The vertical dashed lines indicate the delay times $\textnormal{T}_{\textnormal{f(m)}}$ ($\textnormal{T}_\textnormal{m}$) as the  measured qubit \textit{f(m)} conditionally changes its initial state to the opposite one depending on the fluxon position.}
	\label{fig:proposal}
\end{figure}

\begin{figure}
	\includegraphics[width=1\columnwidth]{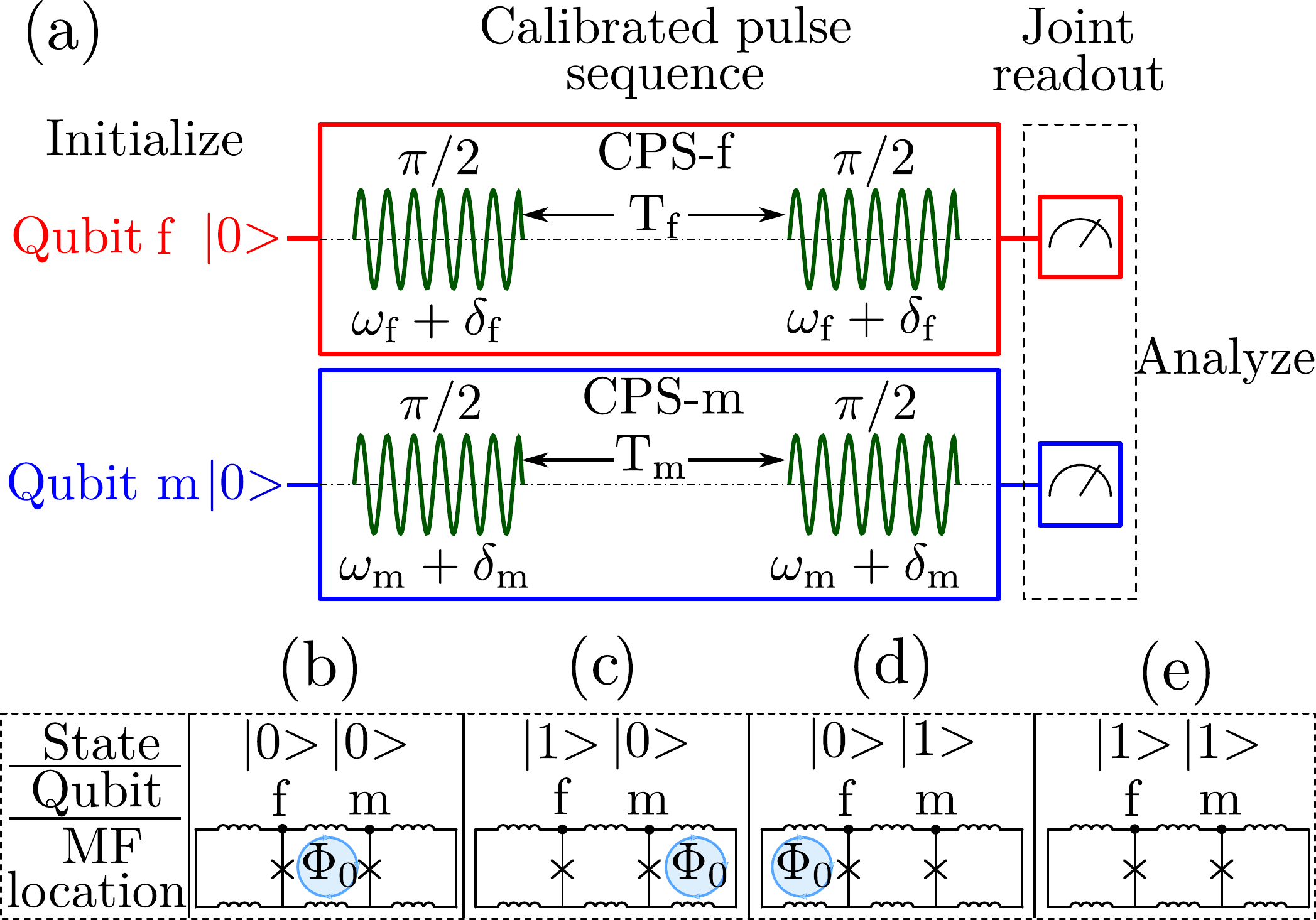}
	\caption{(color online) (a) The proposed measurement protocol for the detection of fluxon position. Firstly, we initialize transmon-qubits (\textit{f} and \textit{m}) to their ground states. Then we apply calibrated pulse sequences (CPS) for the conditional state rotation. The resulting qubits state are then measured by applying a microwave tones to the readout resonators; (b),(c),(d) - Interpretation of measurement results for fluxon located in central (b), right (c) and left (d) cells; (e) - Shown for completeness, there is no fluxon in three-cell SQUID.}
	\label{fig:measurement_protocol}
\end{figure}

The proposed dispersive readout consists of two parts. Firstly, one has to calibrate microwave pulse sequences (CPSs) for the detection of fluxon, see Fig.~\ref{fig:proposal}. This procedure is similar to the two-qubit $CZ$ gate calibration \cite{CZgate, CZgate2}.
There two qubits ZZ interaction leads to a frequency shift of one qubit depending on the state of another. At variance with the $CZ$ gate case, here the transmon qubit (\textit{f},\textit{m}) is coupled with the qutrit. Secondly, obtained CPSs has to be applied for the conditional state rotation of the transmon qubits (\textit{f},\textit{m}), see Fig.~\ref{fig:measurement_protocol}. 
%Secondly, to use the obtained CPSs for the measurements of transmon-qubit states (plasma excitations in Josephson junctions \textit{f(m)}). 

For the calibration procedure we use the Ramsey pulse sequence for transmon-qubit \textit{f} (transmon qubit \textit{m}), i.e. the pre-calibrated $\pi/2$ pulse modulated with frequency $\omega_\textnormal{f}+\delta_\textnormal{f}$ (or $\omega_\textnormal{m}+\delta_\textnormal{m}$), where $\delta_\textnormal{f,m}$ is detuning from the qubit frequency, then another $\pi/2$ pulse applied after a delay time $\Delta\textnormal{t}$, and finally we measure of the transmon-qubit \textit{f} (transmon-qubit \textit{m}) state via coupled resonators. Such Ramsey sequence is presented in Fig.~\ref{fig:proposal}a. Depending on the fluxon location different  Ramsey fringes for transmon-qubit \textit{f} [Fig.~\ref{fig:proposal}b] and qubit \textit{m} [Fig.~\ref{fig:proposal}c] could be observed. Blue lines correspond to the fluxon location in the left cell, while red lines correspond to the fluxon location in the right cell. From Fig. ~\ref{fig:proposal}b one can see that after the particular time delay $\textnormal{T}_\textnormal{f}$ the transmon-qubit \textit{f} should be in the $\ket{0}$ state if fluxon was initially localized in the left or center cell, and in the $\ket{1}$ state if fluxon was initially localized in the right cell due to the additional phase accumulation. The opposite picture is observed in the transmon-qubit \textit{m} state population after the time delay $\textnormal{T}_\textnormal{m}$, see Fig.~\ref{fig:proposal}c. The transmon-qubit \textit{m} should be in the $\ket{0}$ state if fluxon was initially localized in the right cell, and in the $\ket{1}$ state if the fluxon was initially localized in the left or center cell. 
Calibrated time delays are obtained from the condition:
\begin{multline}
	\textnormal{T}_\textnormal{f(m)} = \frac{1}{\delta_{\textnormal{f(m)}}}n= \frac{1}{\delta_{\textnormal{f(m)}}+J^z_{\textnormal{f(m)}}}{(n+1/2)}. \ \ \ \ \ \ \ \ \ \ \
	\label{C1}
\end{multline}
%there $n$ - is the number of perionds of ramsi fringes. Choosing %$\delta_{\textnormal{f}}=2J^x_{\textnormal{f}}$ we define $n=1$ and %
For a particular set of parameters $\delta_{f(m)}=2J^z_{\textnormal{f(m)}}$ and  $J^{\textnormal{z}}_{\textnormal{f}}/h=7.3 \ \si{\mega\Hz}$, $J^{\textnormal{z}}_{\textnormal{m}}/h=6.1 \ \si{\mega\Hz}$) we obtain 
$\textnormal{T}_\textnormal{f}\approx 71 \ ns$ and $\textnormal{T}_\textnormal{m}\approx 83 \ ns$.

With such calibrated CPSs, i.e. the Ramsey pulse sequence with the pre-calibrated  time delays, $T_{\textnormal{f(m)}}$ the 
%The pulse sequences providing the 
plasma modes excitations on \textit{f} and \textit{m} Josephson junctions should swap the ground and excited states of corresponding transmon-qubit. Non-demolition fluxon state measurement protocol, as shown in Fig.~\ref{fig:measurement_protocol}a, thus consist of three steps: (1) - transmon qubits (\textit{f},\textit{m}) initialization in their ground states $\ket{0}$; (2) - application of calibrated pulse sequences (CPS) to realize the conditional swap gates for plasma mode excitations; (3) - transmon-qubit states joint readout and analysis. 

Based on the results of state measurements we can uniquely determine the position of the fluxon in the cells according to the diagram presented in Fig.~\ref{fig:measurement_protocol}b,c,d,e. Thus, the fluxon location in the central cell does not cause an additional phase accumulation in any of the qubits, Fig.~\ref{fig:measurement_protocol}b. As the fluxon is located in the left (right) cell, the CPS displays an additional phase accumulation for the qubit \textit{m} (\textit{f}), as a result, its state will change to the $\ket{1}$, see Fig.~\ref{fig:measurement_protocol}c (Fig.~\ref{fig:measurement_protocol}d). 

\section{Conclusions}
In conclusion, we have numerically and analytically studied the coherent quantum dynamics of a magnetic fluxon trapped in a two- and three-cell SQUIDs (see, Figs.~\ref{fig:two_cells_SQUID} and \ref{fig:full_scematic}). In the range of parameters, as $E_{J\textnormal{f}} \geq E_C$, the coherent tunneling of fluxon between the adjacent cells results in the quantum beats of fluxon. The frequency of such quantum beats is determined by the energy difference between the low-lying levels and rapidly decreases with increasing the Josephson coupling energy $E_J$ (see, Fig.~\ref{fig:7}). In the limit $E_{J\textnormal{f}} \gg E_C$, quantum tunneling of fluxon is strongly suppressed, and the fluxon is localized in one of the cells. In the three-cell SQUID, the fluxon location can be experimentally detected by spectroscopy of plasma mode excitations in Josephson junctions. The quantum dynamics of low-lying energy states is described by the reduced Hamiltonian $\hat{H}_{Qb-Qtr}$ of two transmon-qubits interacting with a qutrit. The experimental protocol allowing one to carry out time resolved measurements of fluxon quantum dynamics is proposed.

\begin{acknowledgments}
 
The authors acknowledge the financial support of the Russian Science Foundation, Project (19-42-04137) and the German Science Foundation (DFG), Grant No. US 18/18-1. 
\end{acknowledgments}

\appendix
\section{\\NUMERICAL METHODS} \label{100}
In this Appendix we describe in detail our numerical approach.
%the solution of  Hamiltonian Eq.\ref{eq3} of the main text.
The energy spectrum and the eigenstates of the Hamiltonian (\ref{eq3}) are obtained from the solution of the time-independent Schr\"odinger equation:
\begin{multline}
\hat{H}\ket{\psi({\varphi}_\alpha|\alpha=\textnormal{f,m})}=E\ket{\psi({\varphi}_\alpha|\alpha=\textnormal{f,m})}, \ \  \ \ \             
\label{eqA1} 
\end{multline}
To simplify the solution we introduce a dimensionless variable for the flux $\hat{\varphi}_\alpha=\frac{2\pi\hat{\Phi}_\alpha}{\Phi_0}$ and canonically conjugate Cooper pair number $\hat{n}_\alpha=\frac{\hat{Q}_\alpha}{2e}$ as well as for external flux variables $\varphi^\textnormal{x}_{1}=\frac{2\pi{\Phi}^\textnormal{x}_{1}}{\Phi_0}$, $\varphi^\textnormal{x}_{2}=\frac{2\pi{\Phi}^\textnormal{x}_{2}}{\Phi_0}$, $\varphi^\textnormal{x}_\textnormal{m}=\frac{2\pi{\Phi}^\textnormal{x}_\textnormal{m}}{\Phi_0}$. 
Writing a momentum-like continuous variable $\hat{n}_\alpha=-{i}\frac{\partial {}}{\partial{\varphi_\alpha}}$  the associated Schr\"odinger equation takes the form of an ordinary differential equation:
\begin{multline}
\hat{H}\left(-{i}\frac{\partial {}}{\partial{\varphi_\alpha}},\hat{\varphi}_\alpha|\alpha=\textnormal{f,m}\right)\ket{\psi}=E\ket{\psi},\  \ \ \ \ \ \ \
\label{eqA2}
\end{multline}
For a two-cell SQUID with a trapped magnetic fluxon as the Eq.\ref{eq4} is applied, the associated Schr\"odinger equation takes a simpler form and it depends only on the variable $\hat{\varphi}_\textnormal{f}$:
\begin{multline}
	\hat{H}_\textnormal{F}\left(-{i}\frac{\partial {}}{\partial{\varphi_\textnormal{f}}},\hat{\varphi}_\textnormal{f}\right)\ket{\psi}=E\ket{\psi},\  \ \ \ \ \ \ \
	\label{eqA3}
\end{multline}
 
This eigenvalue equations were solved by finite-difference methods complemented by exact diagonalization.

For energy levels presented in Fig.~\ref{fig:Full_spectr} with a single fluxon trapped in a single cell, we choose magnetic fields as  $\varphi^\textnormal{x}_{1}+\varphi^\textnormal{x}_{2}=2\pi$, $\varphi^\textnormal{x}_{1}-\varphi^\textnormal{x}_{2} \in[0,4\pi]$, $\varphi^\textnormal{x}_\textnormal{m}=0$. 

Transmon-like transition energies for a system without trapped fluxon were obtained with magnetic fields: $\varphi^\textnormal{x}_{1}=0$, $\varphi^\textnormal{x}_{2}=0$, $\varphi^\textnormal{x}_\textnormal{m}=0$.

\section{Analytical calculations of the quantum beats frequency}\label{200}

Described above numerical method allows us to study energy spectrum in various regimes, including ${\beta} \leq 1$. In the opposite limit, i.e. $\beta \gg 1$, for studying the quantum dynamics of a single magnetic fluxon trapped in a two-cell SQUID we used the quasi-classical description of tunneling through potential barrier developed in \cite{Tunnel_rate}. 
With this approach we obtain the analytical solution for the low-lying energy levels difference, $\Delta (\Phi_\Delta=1.0 \Phi_0)$.
%We considerfluxon tunneling rate Here we assume ${\beta} \geq 1$ and take the %external flux difference ${\Phi}_{\Delta}$ to be close to the flux quantum. 
The quasi-classical approach is valid in the regime as  ${E}_C \ll E_{J\textnormal{f}}$. First of all, we  write the potential energy for our two-cell SQUID:
\begin{multline}
	U=E_{J \textnormal{f}}(1-\cos(\varphi_\textnormal{f})+E_L \left(\varphi_\textnormal{f}-\pi \frac{\Phi_\Delta}{\Phi_0} \right)^2.\ \ \ \  \ \
	\label{eq11}
\end{multline}
%where ${\varphi}_1=\frac{2\pi\hat{\Phi}_1}{\Phi_0}$, %${\varphi}_{\Delta}/2=\frac{{\Phi}_{\Delta}/2}{\Phi_0}$. 
Here, We  eliminate constant term $2\frac{({\Phi}_{\Sigma}/2)^2}{2L}$ for simplicity. As $\Phi_\Delta \simeq 1.0\Phi_0$, extrema of the potential $U(\varphi_\textnormal{f})$  are obtained by solving the transcendental equation:
\begin{multline}
	E_J\sin({{\varphi}_\textnormal{f}})=2E_L{({\varphi}_\textnormal{f}-\pi\Phi_\Delta/\Phi_0)}.\ \ \ \ \ \ \ \ \ \ 
	\label{eq12}
\end{multline}
As $\Phi_\Delta = 1.0\Phi_0$ one can obtain that ${\varphi}_\textnormal{f,max}=\pi$ corresponds to the potential barrier maximum, localized between two minimums ${\varphi}_{{m}R}$ and ${\varphi}_{mL}$. In the limit $\beta \gg 1$:  ${\varphi}_{mR} \approx 2\pi-\pi\frac{2E_L}{E_J}$  and  ${\varphi}_{mL} \approx \pi\frac{2E_L}{E_J}$. 
Neglecting tunneling between the potential wells, we can find energies of two localized states by solving the stationary Schr\"{o}dinger equation for harmonic oscillator:
\begin{multline}
	\left[-4{E}_C\frac{\partial{}^2}{\partial{\varphi_\textnormal{f}}^2} +\frac{V''(\varphi_{m\gamma})}{2}{({\varphi}_\textnormal{f}-\varphi_{m\gamma})^2}\right] \Psi_\gamma({\varphi}_\textnormal{f}) =\\E_\gamma\Psi_\gamma({\varphi}_\textnormal{f}),    \gamma=L,R \label{eq13}.
\end{multline}
Here we used the second order series expansion of the potential energy, $V''(\varphi_{mi})=E_J\cos({\varphi_{m\gamma}})+2E_L$ and $E_\gamma=V(\varphi_{m\gamma})+\frac{\hbar\omega_\gamma}{2}$. The oscillator frequency is:
\begin{multline}
	\hbar\omega_\gamma= 2\pi\sqrt{8E_C(2E_L+E_J\cos{\varphi_{m\gamma}})}. \ \ \ \ \ \ \ \ \
	\label{eq14}
\end{multline}
Now we obtain the amplitude $\Delta$ for quantum tunneling between neighboring wells in the quasi-classical approximation 
%using WKB method 
\cite{Tunnel_rate}. For symmetric wells the amplitude of tunneling $\Delta$ is given by:
\begin{multline}
	\Delta = \frac{\hbar\omega_\gamma}{e\sqrt{\pi}} \exp{(-\int_{\phi_{\gamma1}}^{\phi_{\gamma2}} \sqrt{\frac{1}{E_C}(V(\varphi')-E_\gamma)}d\varphi')}, 
	\label{eq15}
\end{multline}
here boundaries of integration $\phi_{\gamma1,2}$ are the two points at which the potential barrier intersects the energy level: $V(\phi_{\gamma1,2})=E_\gamma$. This analytical dependence is shown on Fig.~\ref{fig:7} (line 2).

For the sake of completeness, we also recall the known asymptotic formula for the splitting in the periodic cosine potential with renormalized coefficients according to \cite{Tunnel_rate2}:
\begin{multline}
	\Delta = 2\sqrt{\frac{2}{\pi}} \sqrt{8\bar E_J\bar E_C} ({\frac{8\bar E_J}{\bar E_C}})^{\frac{1}{4}} \exp{-\sqrt{\frac{\bar E_J}{\bar E_C}}}. \ \ \ \
	\label{eq16}
\end{multline}
where $\bar E_J=E_J[1-\frac{\pi^2}{4\beta}(1-\frac{1}{\beta})]$, $\bar E_C = \frac{E_C}{(1-1/\beta)^2}$.

As a generalization, for studying fluxon dynamic we are only interested in the two lowest eigenstates (fluxon location in right or left potential well), and since the SQUID spectrum is largely anharmonic around the $\Phi_{\Delta}=\pm \Phi_{0}$, and due to Eq.\ref{eq10}, the system Hamiltonian can be reduced to the first two states:
\begin{equation}
	\hat{H}_{\textnormal{2cells-MF}}= \frac{\Delta}{2}\hat{\sigma}_\textnormal{x}+\frac{\epsilon}{2}\hat{\sigma}_\textnormal{z},
	\label{eq17}
\end{equation}
here $\epsilon=\delta \Phi_{\Delta} \frac{\Phi_0}{2 L}$ and $\delta\Phi_{\Delta}=(\Phi_{\Delta}-\Phi_{0})$. This Hamiltonian determines the energy levels splitting as 
$\Delta E_{01}=\sqrt{\Delta^2+\epsilon^2}$.

\end{document}